\documentclass[referee,sn-nature]{sn-jnl}

\usepackage{graphicx}%
\usepackage{multirow}%
\usepackage{amsmath,amssymb,amsfonts}%
\usepackage{amsthm}%
\usepackage{mathrsfs}%
\usepackage[title]{appendix}%
\usepackage{xcolor}%
\usepackage{textcomp}%
\usepackage{manyfoot}%
\usepackage{booktabs}%
\usepackage{algorithm}%
\usepackage{algorithmicx}%
\usepackage{algpseudocode}%
\usepackage{listings}%

\usepackage{physics}
\usepackage{gensymb}
\usepackage{enumitem}

\usepackage{chemformula}
\newcommand{\RR}{\mathbb{R}}
\newcommand{\NN}{\mathcal{N}}
\newcommand{\VV}{\mathcal{V}}

\usepackage{lineno}
\usepackage{etoolbox} 
\newcommand*\linenomathpatch[1]{%
 \cspreto{#1}{\linenomath}%
 \cspreto{#1*}{\linenomath}%
 \csappto{end#1}{\endlinenomath}%
 \csappto{end#1*}{\endlinenomath}%
}
\linenomathpatch{equation}
\linenomathpatch{gather}
\linenomathpatch{multline}
\linenomathpatch{align}
\linenomathpatch{alignat}
\linenomathpatch{flalign}

\begin{document}

\title[Article Title]{Universal wrinkling of 
freestanding 
atomically thin films}

\author[1,6]{\fnm{Jaehyung} \sur{Yu}}
\equalcont{These authors contributed equally to this work.}

\author[2,3]{\fnm{Colin} \sur{Scheibner}}
\equalcont{These authors contributed equally to this work.}

\author[4]{\fnm{Ce} \sur{Liang}}

\author[2,3]{\fnm{Thomas A.} \sur{Witten}}

\author[2,3,5]{\fnm{Vincenzo} \sur{Vitelli}}\email{vitelli@uchicago.edu}

\author[1,3,4]{\fnm{Jiwoong} \sur{Park}}\email{jwpark@uchicago.edu}

\affil[1]{\orgdiv{Department of Chemistry}, \orgname{The University of Chicago}, \orgaddress{\street{5735 S. Ellis Ave.}, \city{Chicago}, \state{Illinois} \postcode{60637}, \country{USA}}}

\affil[2]{\orgdiv{Department of Physics}, \orgname{The University of Chicago}, \orgaddress{\street{5720 S. Ellis Ave.},\city{Chicago}, \state{Illinois}  \postcode{60637}, \country{USA}}}

\affil[3]{\orgdiv{James Franck Institute}, \orgname{The University of Chicago}, \orgaddress{\street{929 E. 57th St.}, \city{Chicago},  \state{Illinois} \postcode{60637}, \country{USA}}}

\affil[4]{\orgdiv{Pritzker School of Molecular Engineering}, \orgname{The University of Chicago}, \orgaddress{\street{5640 S. Ellis Ave.}, \city{Chicago}, \state{Illinois} \postcode{60637}, \country{USA}}}

\affil[5]{\orgdiv{Kadanoff Center for Theoretical Physics}, \orgname{The University of Chicago}, \orgaddress{\street{933 E. 56th St.}, \city{Chicago}, \state{Illinois} \postcode{60637}, \country{USA}}}

\affil[6]{\orgdiv{Current address: Mechanical engineering}, \orgname{Chung-Ang University}, \orgaddress{\street{84, Heukseok-ro, Dongjak-gu}, \city{Seoul}, \postcode{06974}, \country{Republic of Korea}}}

\maketitle

{\bf Atomically thin films, like transition metal dichalcogenides, can now be synthesized at wafer scale, 
achieving the same extreme aspect ratio ($\sim 10^8$) that a sheet of paper would have if it covered an entire city.  
Yet, the intrinsic (i.e. unconfined) three-dimensional shape of these extreme membranes remains a mystery 
because of the very fundamentals of mechanical measurements: to measure such an ultra-thin film, one first needs to simultaneously free it and stabilize it without introducing confining boundaries. 
Here, we introduce a counter-intuitive solution to this problem: place atomically thin films on water. Using atomic force microscopy (AFM) and Raman spectroscopy adapted to water's surface, we reveal that large-scale freestanding membranes spontaneously self-wrinkle
into a universal mechanical state with long
emergent length scales that
follow robust scaling trends. Our analytical and numerical models suggest that these universal trends are controlled by mesoscopic parameters of the polycrystalline domains instead of atomistic details. Moreover, we demonstrate experimentally that the wrinkles result in a large and tunable reduction of elastic stiffness by up to 2 orders of magnitude. The present work 
illuminates the physical properties of the world's thinnest materials at length scales never probed before and highlights their potential for tunable strain-controlled nanomechanical devices. 
}

The mechanics and intrinsic structure of atomically thin materials has remained a topic of intense inquiry and mystery, because of their acute sensitivity to both intrinsic and extrinsic mechanical forces~\cite{Lee2008Measurement,Meyer2007Suspend,Warner2012Science,Bao2009RippleGr}, including thermal fluctuations, structural defects~\cite{Bao2009RippleGr,Kim2018trapOnDrumhead,Fasolino2007intrinsic}, and interactions with their environment~\cite{Lee2013high,Han2019BendGr,Wei2013}. While these features could provide multiple avenues for controlling and patterning mechanical properties for advanced applications~\cite{Kang2015High,Kim2020StochasticJump,Yu2021Perspective,Grosso2015Bending,Blees2015Graphene}, such potential cannot yet be realized.
The difficulty originates from the fundamentals of mechanical measurements: to measure the mechanical state, including the shape and strain distribution, of a thin film, one first needs to simultaneously free it and stabilize it.
However, in atomically thin films, the dominant approach, membrane suspension on a hole, is typically limited to micrometer-scale films due to brittle fracture~\cite{Lee2008Measurement,Lloyd2016bandgap,Kim2018trapOnDrumhead,Brivio2011,Zhang2014FractureTough}. 
Moreover, stabilization inherently biases the membrane's mechanical state. 
For example, drumhead-like suspended atomically thin membranes experience pre-tensions at the solid boundaries (as large as $0.6 N/m$) causing uncontrolled strains~\cite{Lee2008Measurement,Liu2014cNL}, and the process of delaminating a membrane for suspension introduces detrimental perturbations (e.g. tears, strains, and contaminations)~\cite{Ruiz-Vargas2011GrHole,Kim2018trapOnDrumhead}. 
Thus, studies of the mechanical states and properties of atomically thin films are still limited to the micrometer-sized membrane, which is a similar aspect ratio to a letter-sized paper ($10^4$). However, the state-of-the-art synthesis techniques now consistently achieve wafer-scale homogeneous films with an aspect ratio up to $10^8$~\cite{Kang2015High}, the same as a sheet of paper large enough to cover downtown Paris. 
Hence, studying the intrinsic mechanics of large, wafer-scale atomically thin membranes necessitates an experimental platform with a seemingly impossible condition: freeing and holding large-scale membranes with almost no perturbations to their mechanical states. 

\begin{figure*}[t!]
\centering
\includegraphics[width=0.8\textwidth]{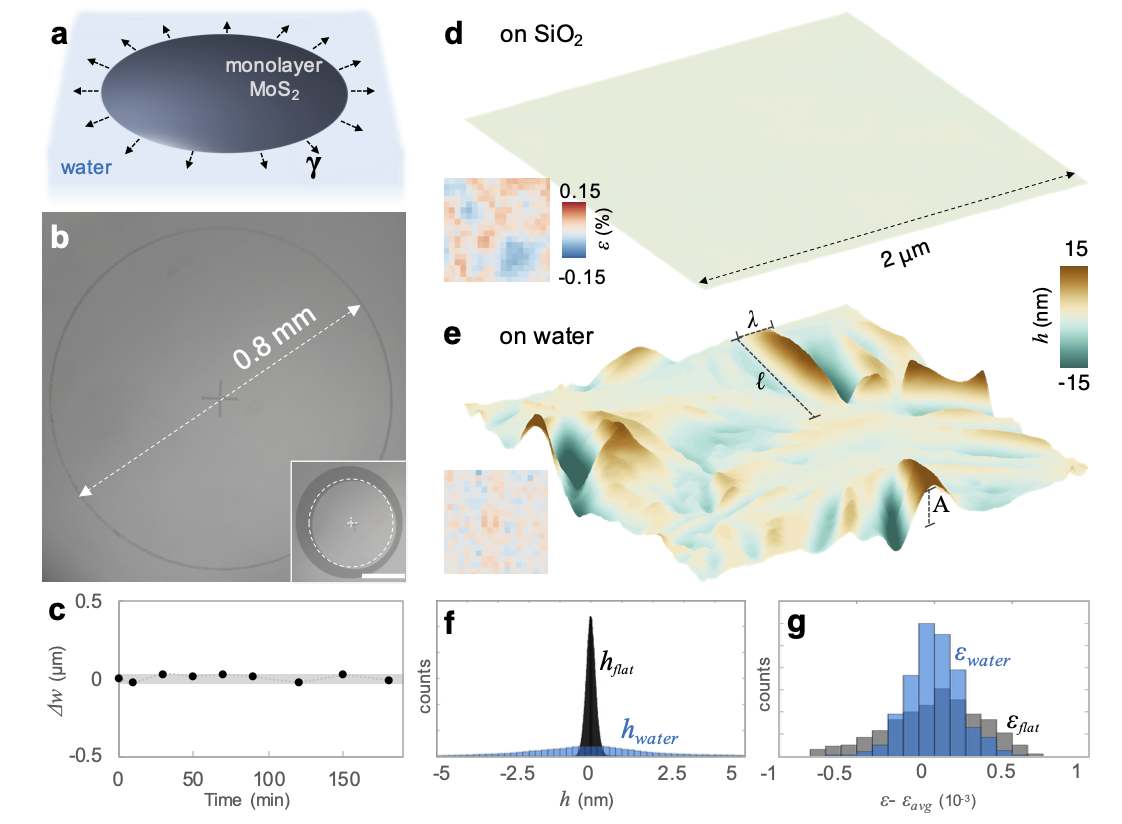}
 \caption{{\bf Freestanding atomically thin \ch{MoS2} on water.}~~{\bf a.}~A schematic of an atomically thin membrane on water experiencing surface tension $\gamma$ at its boundary. 
 {\bf b.}~An optical micrograph of a mm-scale floating \ch{MoS2} membrane with a gap of width $w \leq 2 \mu m$  created by laser patterning. (inset) A membrane of the same dimensions imaged 2 hours after patterning a wider gap ($w \approx 100 \mu m$). The dotted line depicts the initial position of the membrane. Scale bar = $400~\mu m$. {\bf c.}~The time-dependent change of the gap distance between floating and anchored \ch{MoS2} with a narrow gap ($w \leq 2 \mu m$). The standard deviation is shown in grey.  
 {\bf d-e.}~AFM height and strain maps (inset) of monolayer \ch{MoS2}  (d) as grown on \ch{SiO2}-Si substrate and (e) on water. The symbols $A$, $\lambda$, $\ell$ denote the average height, width and length of the wrinkles, respectively.
 {\bf f-g.}~The histogram of (f) height and (g) strain for monolayer \ch{MoS2} on \ch{SiO2}-Si (grey) and on water (blue).
  }
  \label{fig:fig1}
\end{figure*}

Here, we present a counterintuitive solution: use water as a substrate (Fig.~\ref{fig:fig1}a). 
In this case, the membrane is held by the weak, spatially-uniform surface tension of water ($\gamma\approx70~mN/m$), which causes extremely small strains for most materials. By direct delamination~\cite{Yu2022NL}, pristine membranes that are several orders of magnitude larger (up to 10 cm) than their suspended counterparts~\cite{Lee2008Measurement,Ruiz-Vargas2011GrHole,Lee2013high} can be placed directly on water. \textcolor{black}{Our solution allows us to directly image the 3D shape $h(x,y)$ of free-floating atomically thin films, as well as 
key mechanical properties, including their strain field $\varepsilon(x,y)$ and force response $F(x,y)$ to point indentations.}
As described below, we apply this method to a representative wafer-scale 2D material, monolayer \ch{MoS2}. We discover that---in the absence of tethered boundaries---the monolayer exhibits large-amplitude wrinkles. We reveal that the wrinkle geometry obeys universal scaling relations \textcolor{black}{that provide a path towards 3D shape control without atom-level information. Moreover the wrinkles dramatically alter} the effective mechanical properties of the unconfined membrane. 
Taken together, our findings provide guidelines for  
controlling the 3D shape and the mechanical strength of atomically thin membranes in mechanically free environments.

\begin{center}
{\bf Realizing freestanding atomically thin films on water}
\end{center}
Figure~\ref{fig:fig1}b shows optical micrographs of mm-scale monolayer \ch{MoS2} (brighter) on water (darker). In each image, a circular \ch{MoS2} disk is freely floating, separated from the outer \ch{MoS2} region by a small gap (width $w$). The \ch{MoS2} disks are produced in three steps: (i) a wafer-scale continuous \ch{MoS2} monolayer, synthesized using metal-organic chemical vapor deposition~\cite{Kang2015High}, is delaminated on water in a semi-closed container with a lid; (ii) the monolayer is brought into contact with the lid (with openings for measurements) by slowly raising the water level (See Supplementary Fig.~\ref{Sfig:container}); (iii) a millimeter sized raft (or disk) is liberated within an opening by cutting a line of gap using a focused laser beam~\cite{Yu2022NL,Poddar2022NL}. At this stage, the interior disk is freely floating, primarily experiencing the surface tension of water ($\gamma$) at its edge and buoyancy force from beneath. The latter is negligible for nanoscale membrane deformations that are far smaller than the capillary length $\ell_C \sim 3~mm$. 

\ch{MoS2} membranes floating on water are effectively immobilized from the lateral motion by the large membrane size and by membrane edge repulsion.
On long time scales ($\sim$hr), we find that a mm-sized disk surrounded by an open water surface (Fig.~\ref{fig:fig1}b, inset; $w \sim 100~\mu m$) slowly drifts, but such slow displacement is prevented when we narrow the gap (Fig.~\ref{fig:fig1}b, main panel; $w \sim 2~\mu m$). Time-lapse images confirm that the lateral translocation for the narrow gap is negligible, with the gap-width variation $\Delta w$ measured to be less than the instrument resolution ($30~nm$) for 3 hrs (Fig.~\ref{fig:fig1}c). During this time, the inner disk never touches the outer \ch{MoS2} (see Supplementary Fig.~\ref{Sfig:drift}). This suggests that there exists a repulsive interaction between the adjacent \ch{MoS2} edges, which we ascribe to the electrostatic repulsion due to charges in the membrane~\cite{Yu2022NL}. In addition, the short time-scale jitter observed for micron-scale membranes~\cite{Yu2022NL} is negligible for a mm-sized membranes, 
likely due to the increased hydrodynamic drag 
on the larger \ch{MoS2}-water interface.

\begin{center}
   {\bf Observation of wrinkles in atomically thin films}
\end{center}

These immobilized large-scale \ch{MoS2} membranes on a static water surface can then be moved from one instrument to another for high resolution imaging of their mechanical states. Figures~\ref{fig:fig1}d-e show a representative example; we measure height maps [$h(x,y)$; main panels] constructed via atomic force microscopy (AFM) and strain maps [$\varepsilon{(x,y)}$, insets] taken using Raman spectroscopy~\cite{Conley2013NL,Michail2016APL,Kukucska2017PSSB} for the same \ch{MoS2} monolayer as grown on a \ch{SiO2}-Si substrate (Fig.~\ref{fig:fig1}d) and after it is placed on water (Fig.~\ref{fig:fig1}e). For AFM measurements on water, we optimize the attractive tapping mode of AFM imaging, which becomes fully compatible with membranes on water by minimizing the perturbation to their mechanical states (See Supplementary Fig.~\ref{Sfig:afm}). While \ch{MoS2} appears uniformly flat on its as-grown substrate with the same flatness as \ch{SiO2}-Si~\cite{Kang2017}, the exact same sheet of \ch{MoS2} on water is far from flat. Instead, the membrane on water forms wrinkles comprising micron-scale ridges and valleys with a valley-to-peak amplitude up to 30 nm, tens of times larger than the membrane thickness ($\sim 0.6 ~nm$)~\cite{Mak2010MoS2}. Multiple AFM images taken from the region yield identical results, confirming (i) that these wrinkles are static and athermal and (ii) that the stability determined by monitoring the gap in Fig.~\ref{fig:fig1}b-c translates into measurement reproducibility. The in-plane strain maps of the membrane show similarly striking difference on the two substrates. 
On \ch{SiO2}-Si, we observe large, spatially heterogeneous regions (each micron-scale) with negative (blue) or positive (red) in-plane strain (difference $\simeq 0.3\%$). The strain then diminishes everywhere when the membrane is on water. These results, combined with the histograms in Fig.~\ref{fig:fig1}f and Fig.~\ref{fig:fig1}g, present our first main finding: a large-scale \ch{MoS2} monolayer spontaneously wrinkles in its free state on water and releases the in-plane strain that was present in its substrate-confined flat state. 
This is an intrinsic property of the \ch{MoS2} membrane, as it is not tethered at its boundary and the surface tension of water can cause only negligible strain ($\sim0.03\%$).	
Furthermore, a crucial distinction between the wrinkles we observe and those in macroscopic thin sheets on water~\cite{Pocivavsek2008Stress,Huang2007Capil,Tobasco2022Exact,Aharoni2017smectic} is that gravity is negligible for deformations in our system, because both the elasto-gravitational length $\ell_g \approx 4~ \mu m$ and capillary length $\ell_C \approx 3~mm $ 
induced by water are larger than the deformation wavelengths we observe ($\lesssim 1~\mu m$).

\begin{figure*}[t!]
\centering
\includegraphics[width=1.0\textwidth]{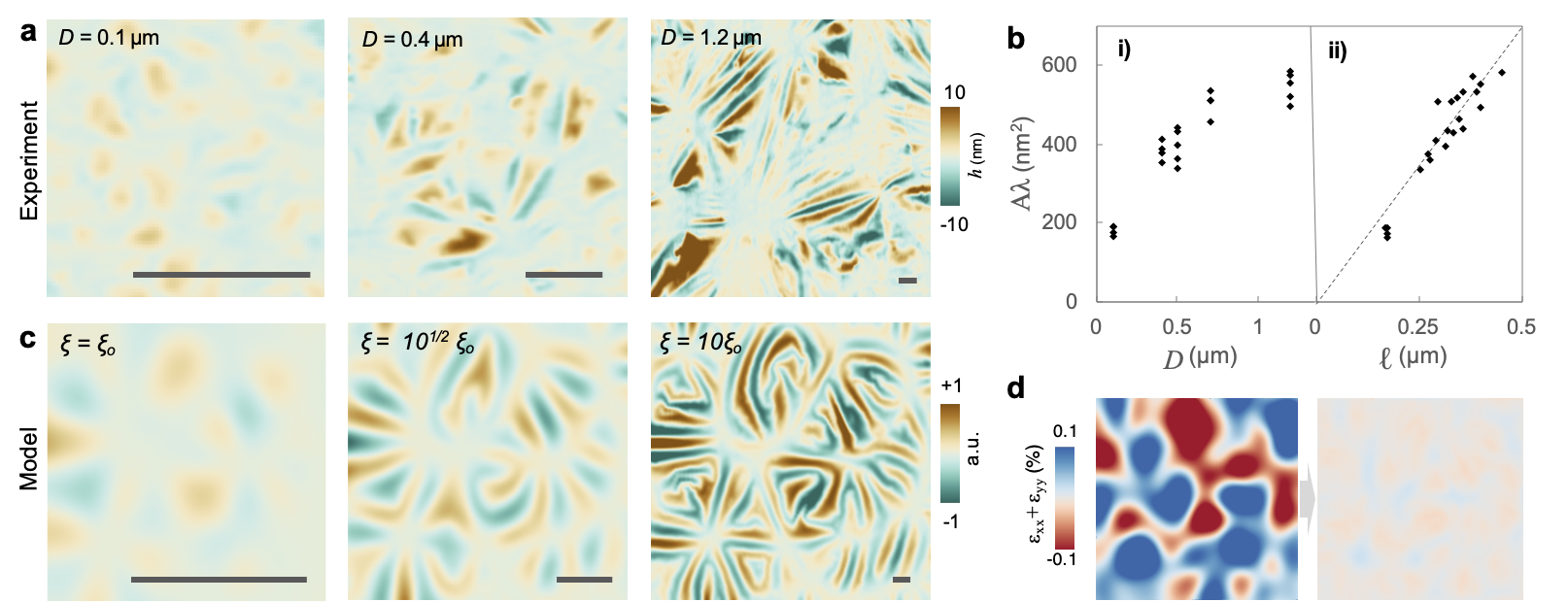}
  \caption{{\bf The universal scaling of wrinkle morphology.}~~{\bf a.}~Experimental measurements of \ch{MoS2} surface topography on water with the average crystalline domain size $D$ increasing  from left to right. Scale bar = 500 nm. 
  {\bf b.}~The product of the average wrinkle width $\lambda$ and amplitude $A$ is plotted as a function of~i)~average domain size $D$ and~ii)~wrinkle length $\ell$. The theoretical scaling in Eqs.~(\ref{eq:scale1}-\ref{eq:scale3}) predicts a linear relationship $A \lambda \propto t \ell$, where $t$ is the elastic membrane thickness.  
  {\bf c.}~The height fields resulting from the numerical relaxation of a continuum model in which a thin membrane is subjected to the random strains in Eq.~(\ref{eq:distribution}). The correlation length $\xi$, a proxy for domain size $D$, increases from left to right. The strain variation $\Delta \varepsilon$ is held constant.  
  {\bf d.}~(left)~The trace of the strain tensor in Eq.~(\ref{eq:distribution}). (right)~The trace of the strain tensor after three-dimensional relaxation.}
  \label{fig:fig2}
\end{figure*}

This observation raises three fundamental questions: the relation between the flat-state strain distribution [$\varepsilon_\text{flat}(x,y)$] and the wrinkle morphology on water [$h_\text{water}(x,y)$], the potential universality of the wrinkling, and the origin of $\varepsilon_\text{flat}(x,y)$. To answer these, we repeat the experiments shown in Fig.~\ref{fig:fig1} for multiple \ch{MoS2} membranes that are nearly identical except for the average domain size, $D$, measured using the AFM phase images (Supplementary Fig.~\ref{Sfig:heightphase}). Figure~\ref{fig:fig2}a shows experimental $h_\text{water}(x,y)$ maps taken from three representative \ch{MoS2} membranes with different $D$ = 0.1, 0.4 and 1.2 $\mu m$ (see more data in Supplementary Figs.~\ref{Sfig:heightphase} and \ref{Sfig:grainsize}). We find that the wrinkle morphology evolves systematically with $D$. First, the wrinkle amplitude, $A$, increases with $D$. Second, both the length, $\ell$, and width, $\lambda$, of a wrinkle increase with $D$. Third, the typical wrinkle shape evolves from being isotropic ($D = 0.1~\mu m$; $\ell/\lambda \sim 1$) to more elongated ($D = 1.2~\mu m$; $\ell/\lambda \gg 1$). For larger $D$, a wrinkle can even span several domains along the length, while multiple wrinkles can be found within a single domain along the width. We summarize our data from many ($\sim$20) AFM images in Fig.~\ref{fig:fig2}b~i), where we plot the product $A\lambda$ versus $D$, revealing a data collapse and a monotonically increasing
relationship.

\begin{center}
    {\bf Continuum model and universal scaling}
\end{center}
To understand the mechanical origin of the morphological changes, we notice that the strain field $\varepsilon_\text{flat} (x,y)$ is highly disordered (see Fig.~\ref{fig:fig1}d). To 
interrogate how this property affects wrinkling,
we consider a minimal, continuum model in which the strains are disordered with typical variation $\Delta \varepsilon$ and correlation length $\xi$. Specifically, the strain field we apply is (See an example in Fig.~\ref{fig:fig2}d, left):
\begin{align}
    \varepsilon_{ij} (\vb x) = \frac{\Delta \varepsilon}{4 \pi \xi } \int  e^{- \frac{\abs{\vb x - \vb y}^2}{2 \xi^2 } } \eta_{ij} (\vb x - \vb y) \dd^2 y,\label{eq:distribution}
\end{align}
where  $\eta_{ij}(\vb x)$ is a field of identical and independently distributed random matrices (see Supplementary Information). 
After applying this initial strain, we allow the membrane to relax in three dimensions (Fig.~\ref{fig:fig2}d, right) with the goal of minimizing its bending and stretching energy. Figure~\ref{fig:fig2}c shows three height maps, $h(x,y)$, after numerical relaxations with increasing $\xi$. 
The model is entirely continuum and does not need atomistic details of a specific crystalline structure. Nonetheless, we expect $\xi$ in the model to be a proxy for the domain size $D$ in the experiments.
While details of the wrinkle shapes differ between the abstract model and \ch{MoS2}, the shared evolution of their morphologies highlights the key role of the strain correlation length $\xi$. It also suggests that the observed wrinkling is not particular to \ch{MoS2} on water; instead, it is a general feature of unconfined 2D media with disordered strains.

Since the behavior in Fig.~\ref{fig:fig2}c is not contingent on atomistic details of \ch{MoS2}, quantitative features of the wrinkles can be rationalized via continuum elasticity theory. In absence of a substrate, wrinkle geometry can be described by tension (or curvature) induced wrinkling~\cite{Cerda2003Geometry,Davidovitch2011Prototypical,Vella2018WaterIndent}, in which local compression causes wrinkles, and transverse tension determines their amplitude and wavelength.
Here we apply this principle to disordered strains, in which the compression and tension arise due to geometrical mismatch among domains. 
We expect the typical compression to be proportional to 
the strain variation
$\Delta \varepsilon$; the typical tension to be proportional to the strain variation times the Young's modulus $Y \Delta \varepsilon$; and the typical tensile persistence length to be proportional to $\xi$. Taken together, the scaling law of wrinkling geometry becomes (see Methods): 
\begin{align} 
A \propto& \, \qty(\Delta \varepsilon)^{1/4} \xi^{1/2} t^{1/2} \label{eq:scale1}  \\  
\lambda \propto& \, \qty(\Delta \varepsilon)^{-1/4} \xi^{1/2} t^{1/2} 
\label{eq:scale2}  \\
\ell \propto& \, \xi  \label{eq:scale3}
\end{align} 
where $t\approx \sqrt{ 12 (1-\nu^2) B/Y}$ is the elastic thickness of a membrane with bending modulus $B$, Young's modulus $Y$, and Poisson's ratio $\nu$. Equations~(\ref{eq:scale1}-\ref{eq:scale3}) make testable predictions. As domain size $D$ (and hence $\xi$) increases, all three dimensions (amplitude, width, and length) of the wrinkles grow, but the length $\ell$ grows the fastest, causing an elongation of the wrinkles as observed in experiments. 
For a quantitative comparison, we combine Eqs.~(\ref{eq:scale1}-\ref{eq:scale3}) to obtain a scaling $A \lambda \propto t \ell$ that is independent of $\Delta \varepsilon$. In the Fig.~\ref{fig:fig2}b~ii), we plot $A \lambda$ as a function of $\ell$, revealing a linear relationship consistent with the scalings in Eqs.~(\ref{eq:scale1}-\ref{eq:scale3}). 
The scaling $A \lambda \propto t \ell$ predicts that the slope of the fitted line ($m$) should be proportional to the elastic thickness ($t$) of the membrane and therefore of a similar order of magnitude. For monolayer \ch{MoS2}, the elastic thickness is $t\approx 0.3$ nm and the slope measured from experiments is $m\approx 1.0 $ nm, consistent with expectations.

\begin{figure*}[t!]
    \centering
    \includegraphics[width=1.0\textwidth]{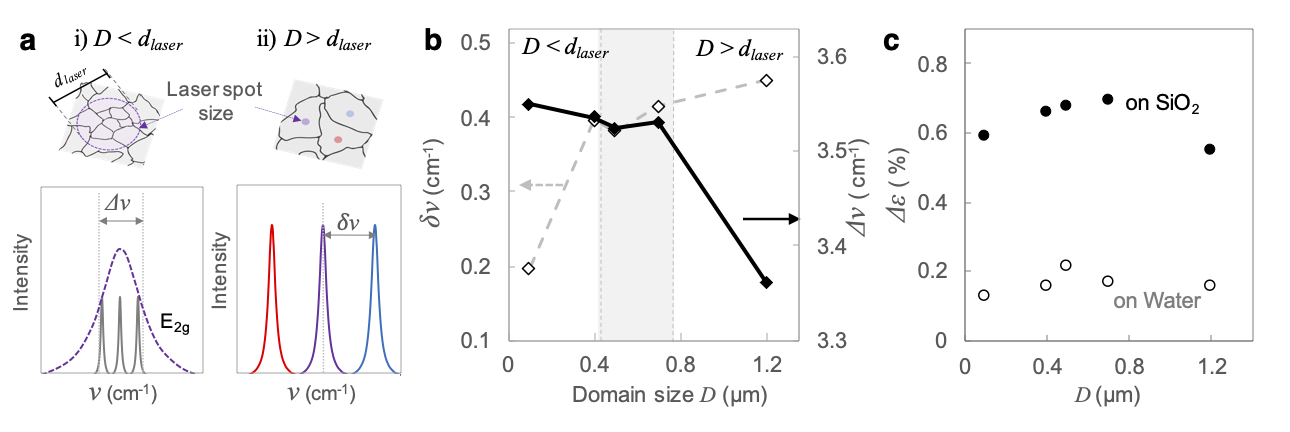}
    \caption{{\bf The domain-size-dependent strain distribution of polycrystalline \ch{MoS2}.}~~{\bf a.}~(top)~The AFM-characterized grain boundary structures of \ch{MoS2} are shown. The colored circles illustrate the relative size of the Raman laser spot for i) small domains ($D<d_{laser}$) and ii) large domains ($D>d_{laser}$). (bottom)~The corresponding Raman $E_{2g}$ peaks are shown. 
    {\bf b.}~Comparison of standard deviation ($\delta v$, left) and average FWHM  ($\Delta v$, right) of the $E_{2g}$ peak in Raman maps for \ch{MoS2} with varying domain sizes.  
    {\bf c.}~The inferred total strain variation $\Delta \varepsilon$ as a function of grain size $D$. 
    }
    \label{fig:fig3}
\end{figure*}

\begin{center}
{\bf Probing strain distribution}
\end{center}
The above analysis implicitly assumes that the strain field originates from the presence of grain boundaries. 
We directly probe this relationship using Raman spectroscopy. First,
Raman spectra taken during growth reveal that 
the $\varepsilon_\text{flat}(x,y)$ first emerges when isolated domains merge to form a continuously connected sheet; if the growth is stopped before the domains connect, we observe much smaller $\Delta\varepsilon$ (see Supplementary Fig.~\ref{Sfig:formation}). Second, to directly discern the relationship between grain size and effective parameters, such as  $\xi$ and $\Delta \varepsilon$, we consider two crucial features of the Raman mapping: the $E_{2g}$ peak center $v$ and peak width $\Delta v$, which broadens due to the convolution of multiple strain states, as conveyed schematically in Fig.~\ref{fig:fig3}a.  
For larger $D$ ($1.2 \mu m > d_{laser}$, laser spot size) in Fig.~\ref{fig:fig3}b, each Raman spectrum corresponds to the strain state of an individual domain (Fig.~\ref{fig:fig3}a, ii). We observe that $v$ fluctuates over the micron scale while $\Delta v$ remains uniformly small, closer to the width $\Delta v_0$ from exfoliated \ch{MoS2}. For smaller $D$ ($0.1 \mu m < d_{laser}$) in Fig.~\ref{fig:fig3}b, by contrast, each spectrum measures multiple domains (Fig.~\ref{fig:fig3}a, i), and both $v$ and $\Delta v$ are spatially uniform, while $\Delta v$ becomes significantly larger due to inhomogeneous broadening. 
Together, these trends support that the origin of $\varepsilon_\text{flat}(x,y)$ is incompatible strains trapped within the domains, and the strain correlation $\xi$ depends strongly on the domain size $D$.

Likewise, we use the approach above to estimate the magnitude of the strain variation $\Delta \varepsilon$ as a function of $D$ (Fig.~\ref{fig:fig3}c). We utilize the convolutional relationship $(\Delta \varepsilon/\alpha)^2 = (\delta v)^2+ (\Delta v)^2-(\Delta v_o)^2$, where $\delta v$ is the standard deviation of $v$, and $\alpha$ is the Raman $E_{2g}$-to-strain coefficient~\cite{Conley2013NL}. The results are shown for \ch{MoS2} on \ch{SiO2}-Si (solid) and water (empty).
We find that $\Delta \varepsilon_\text{flat}$ remains large ($0.5-0.7 \%$) on \ch{SiO2}-Si for all $D$. The lack of a clear dependence on $D$ is consistent with the interpretation that the change in wrinkle morphology shown in Fig.~\ref{fig:fig2}a is primarily driven by $\xi$, which increases systematically with $D$. Moreover, $\Delta \varepsilon_\text{water}$, the strain variation on water is much smaller ($\lesssim 0.2 \%$). This reduction  further confirms that the membrane releases most of its strain energy ($\sim 90\%$) upon wrinkling. We note that the strain after relaxation is likely to yield \ch{MoS2} monolayers with uniform electrical and optical properties as the band gap of monolayer \ch{MoS2} is known to change by approximately 0.1 eV for $1\%$ of strain~\cite{Lloyd2016bandgap}.

\begin{figure*}[t!]
    \centering
    \includegraphics[width=1.0\textwidth]{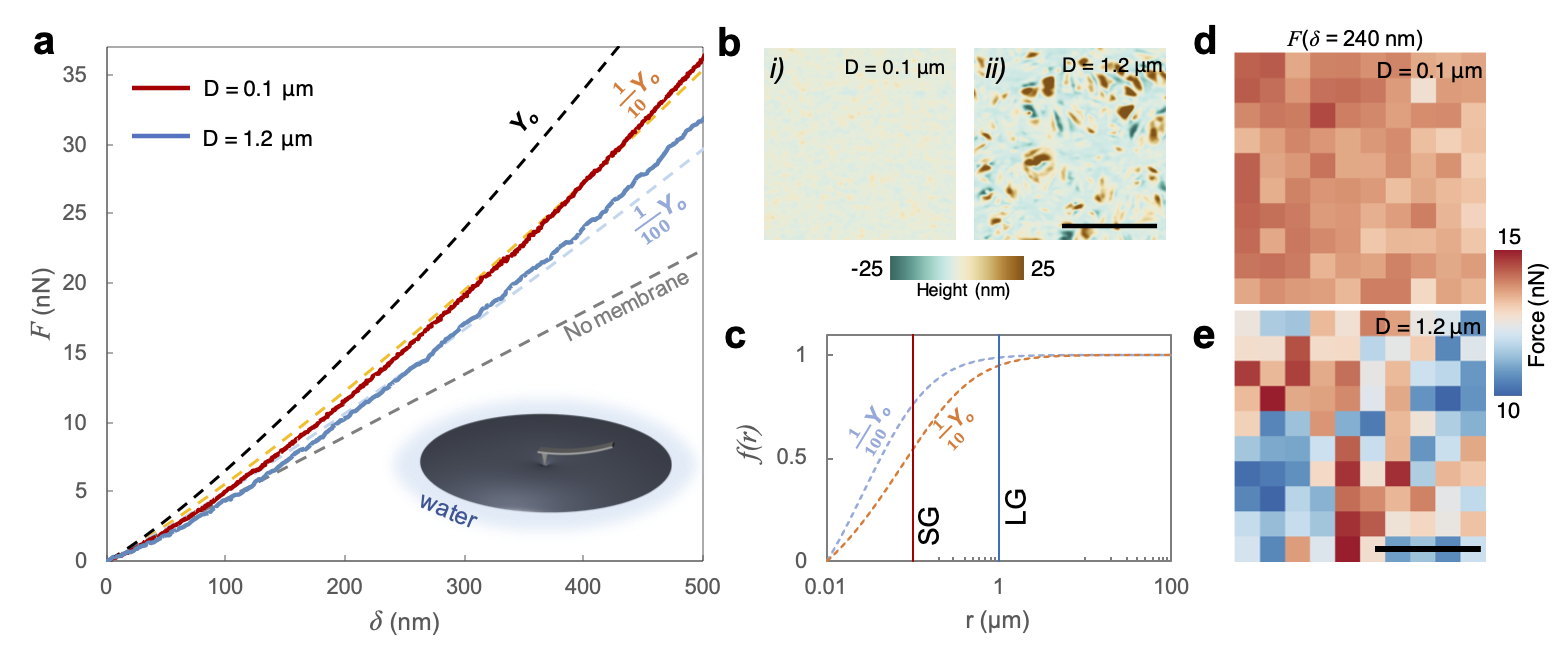}
    \caption{{\bf The effective stiffness and heterogeneity of wrinkled membranes.}~~{\bf a.}~Force-displacement curves measured by AFM indentation of monolayer \ch{MoS2} on water. Solid red and blue curves show the measured $F(\delta)$ for small-grain ($D = 0.1~\mu m$) and large-grain ($D = 1.2~\mu m$) \ch{MoS2}, respectively. Dashed theoretical curves show the response of membranes with effective Young’s moduli $Y_\text{eff} =Y_0$ (black), $Y_\text{eff} = Y_0/10$ (red), and $Y_\text{eff} = Y_0/100$ (blue), where $Y_0 =1.7 \times 10^2~Nm$ is the Young’s modulus of flat, single-crystal \ch{MoS2}. The gray dashed line corresponds to water's surface without a membrane. (inset) Schematic of measurement.  
    {\bf b.}~The surface topography of floating \ch{MoS2} with domain sizes i) $D=0.1~\mu m$ and ii) $D=1.2~\mu m$. Scale bar = $5~\mu m$. 
    {\bf c.}~Plots of $f(r)$, the fraction of strain energy stored within radius $r$ [see Eq.~(\ref{eq:fraction})], for $Y_\text{eff} = Y_0/10$ (red) and $Y_\text{eff} = Y_0/100$ (blue)
    and indentation depth $\delta =240~nm$. 
    The vertical dashed lines denote average domain size $D = 1.2~\mu m$ (LG), and $D = 0.1~\mu m$ (SG). For $D=0.1~\mu m$, any individual grain contains a small fraction of the cumulative energy. For $D=1.2~\mu m$, a single grain contains the vast majority of the strain energy.  
    {\bf d-e.}~The spatially resolved force response map for (d) $D=0.1~\mu m$ and (e) $D=1.2~\mu m$ on the areas presented in (b). The force is measured at the indentation depth $\delta = 240~nm$. Scale bar = $5~\mu m$. }
    \label{fig:fig4}
\end{figure*}

\begin{center}
    {\bf Wrinkle-induced softening and mechanical heterogeneity}
\end{center}
Using water as a substrate, we experimentally determine how wrinkles affect the average and spatially resolved force  response [$F(x,y)$] of \ch{MoS2} to point indentations. 
Figure~\ref{fig:fig4}a shows representative force-response curves (force $F$ vs indentation depth $\delta$) measured using AFM indentation from two \ch{MoS2} membranes with different $D$ ($0.1~\mu m$, red; $1.2~\mu m$, blue). To contextualize our measurements, we numerically solved the radial F\"oppl-von K\'arm\'an equations for a circular membrane (see Supplementary Information) with an effective Young's modulus $Y_\text{eff}$. Figure~\ref{fig:fig4}a contains three curves calculated for $Y_\text{eff} = Y_0$, $Y_0/10$, and $Y_0/100$ (black, red, and blue dashed lines), where $Y_0 = 1.7 \times 10^2~Nm$ is the Young’s modulus of single-crystal \ch{MoS2}~\cite{Box2017waterIndent,Vella2018WaterIndent,Liu2014cNL}. The solutions indicate that $F(\delta)$ for a membrane on water increases relative to bare water (dashed grey line) with greater stiffening for larger $Y_\text{eff}$. 
Comparing the experimental data and the model allows us to estimate $Y_\text{eff}$ for the wrinkled membranes. This comparison constitutes our third main finding: the wrinkles soften the Young’s modulus by up to 2 orders of magnitude. 
Perhaps counter-intuitively, as the domain size $D$ becomes larger, $Y_\text{eff}$ of wrinkled \ch{MoS2} moves further from that of single-crystal \ch{MoS2}. This trend is consistent with the scaling of the wrinkle morphology in Fig.~\ref{fig:fig2}: as $D$ increases, the wrinkles become taller ($A \propto \xi^{1/2}$), enhancing the accordion effect in which the flattening of wrinkles reduces the effective Young’s modulus~\cite{Min2011,Ruiz-Vargas2011GrHole,Kosmrlj2013mechanical}.

However, the effective medium picture alone does not capture the possible effects of spatial heterogeneity. Upon indentation, the response is determined by the local mechanical properties of the membrane within a radius 
\begin{equation}
r_\text{res} \propto  \sqrt{\frac{Y_\text{eff}}{2\gamma}} \delta 
\end{equation}
which contains most ($> 90\%$) of the membrane elastic energy (see Fig.~\ref{fig:fig4}c and Supplementary Information).  This crucial length scale, $1 - 2~\mu m$ in our experiments, sets the effective spatial resolution of mapping based on indentation point probes. Therefore, the large diameter ($\sim 2~mm$) of our membranes on water relative to this micron-scale resolution enables the mapping of the force response $F(x,y)$ to deep AFM indentation. In Fig.~\ref{fig:fig4}d-e, we present $F(x,y)$ maps at $\delta = 240~nm$ measured from two \ch{MoS2} membranes with $D = 0.1 ~\mu m$ and $1.2~\mu m$, respectively (height maps shown in Fig.~\ref{fig:fig4}b). The force response from the small domain \ch{MoS2} is larger (by up to $50\%$) in magnitude than for large domain \ch{MoS2}, consistent with the $Y_\text{eff}$ picture in Fig.~\ref{fig:fig4}a. Furthermore, the response of the small domain \ch{MoS2} is more spatially homogeneous than the response of larger domains. The noticeable difference in homogeneity reflects the crossover between the domain size and mapping resolution: For $D < r_\text{res}$, many wrinkles and domains participate in a single indentation, whereas for $D > r_\text{res}$, only a small number of wrinkles and domains are within $r_\text{res}$.
This crossover also suggests that similar force and tear strength measurements can be conducted reproducibly as long as the points of measurements or local tears are separated by more than $r_\text{res}$.

\begin{center}
    {\bf Outlook}
\end{center}
By harnessing water, an everyday material, as a tether-free measurement platform, we have unveiled the universal wrinkling of atomically thin films, which is key to unlock their advanced functionalities. Our results show that 
coarse-grained quantities, such as the strain variation $\Delta \epsilon$ and the correlation length $\xi$, can serve as mesoscopic control parameters that are experimentally measurable and not contingent on atomistic details. 
This opens up future possibilities, such as
strain controlled optoelectronic membranes and flexomagnetic materials that couple light-matter interactions and magnetism with strain in the atomically thin limit.

\clearpage 

\clearpage

\clearpage

\clearpage

\clearpage

\backmatter

\bmhead{Acknowledgments}

This work is primarily funded by the University of Chicago MRSEC, supported by the NSF under Award Number DMR-2011854.
J.P., J.Y., and C.L acknowledge partial support from the Cornell Center for Materials Research with funding from the NSF MRSEC program (DMR-1719875) as well as the Air Force Office of Scientific Research projects (FA9550- 21-1-0323 and FA9550-18-1-0480).
C.S.~acknowledges support from the Bloomenthal Fellowship and the National Science Foundation Graduate Research Fellowship under Grant No.~1746045. V.V.~acknowledges support from the Simons Foundation, the Complex Dynamics and Systems Program of the Army Research Office under grant W911NF-19-1-0268.

\bmhead{Declarations} Not applicable.

\clearpage

\renewcommand{\theequation}{M\arabic{equation}}
\setcounter{equation}{0}
\renewcommand{\thesubsection}{\arabic{subsection}.}
\setcounter{subsection}{0}
\section*{Methods}
\subsection{Fabrication of floating monolayer \ch{MoS2} films}

The monolayer \ch{MoS2} is synthesized via metal-organic chemical vapor deposition (MOCVD) on \ch{SiO2}-Si substrate~\cite{Kang2015High}. Prior to floating the membrane, the grain size of the monolayer \ch{MoS2} membrane is characterized using the phase image obtained from non-contact AFM scans (refer to Supplementary Fig.~\ref{Sfig:heightphase}). 
Then, the as-grown \ch{MoS2} is placed 45$\degree$ tilted in the Teflon bath. The water is introduced in the bath by using a syringe pump, gradually raising the water surface level at a rate of 1 $mm/hr$~\cite{Yu2022NL}. 
As the water level increases, water molecules intercalate between the monolayer \ch{MoS2} and the substrate, allowing the membrane to float without the need for solvents or polymers.  The floating \ch{MoS2} membrane is delicately retrieved using a small, potable water bath [size of 25mm (W) x 15 mm (L) x 4 mm (H)] with the metal grid (Nickel-Brass shadow mask, Photo-sciences) on the top (See Supplementary Fig.~\ref{Sfig:container}). The grid contains holes of varying sizes, ranging from 200 $\mu m$ to 2 $mm$. Subsequently, high-power, 532 nm laser patterning is employed to cut the floating \ch{MoS2} membrane into dimensions that fit the size of the holes, thereby preventing any unnecessary pinning of the membrane to the metal grid.

\subsection{Characterization of membrane topography and strain distribution}

The surface topography of monolayer \ch{MoS2} on water is characterized by non-contact mode with Assylum MFP-3D BIO.  To minimize perturbations caused by the AFM probes, we use a force modulation AFM probe (OPUS 240NC-AG) and employ a soft-landing procedure while approaching the tip. The surface topography is measured in attractive mode scans (phase $>90\degree$) to minimize the force exerted on the surface during the surface topography measurement. We verify that the repulsive mode scan disrupts the surface topography (See Supplementary Fig.~\ref{Sfig:afm}), whereas the attractive mode scan consistently reproduces the same surface topographies across multiple scans.
The strain distribution is characterized using Raman spectroscopy with the HORIBA LabRAM HR Evolution Confocal Raman Microscope. To generate the strain map, a 532 nm laser is used to scan the $10 \mu m \times 10 \mu m$ area of the \ch{MoS2} membrane in a point-by-point manner. From the Raman spectrum, the in-plane strain is determined by measuring the positions of the $E_{2g}$ and $A_{1g}$ peaks through Lorentzian fitting. The calculated peak locations are then employed to determine the in-plane strain, following the methodology outlined in the literature~\cite{Michail2016APL}.

\subsection{Force-indentation curve of floating monolayer \ch{MoS2}}

The force-indentation measurements were conducted using the AFM (MFP-3D BIO) equipped with the Force-modulated probe (OPUS 240NC-AG). To ensure accurate force measurements on water, the spring constant of the AFM probe was calibrated using thermal tuning. Additionally, the initial parameters were adjusted by applying pressure to the \ch{MoS2} sample on a hard substrate (e.g., \ch{SiO2}-Si). Prior to extracting each indentation curve, the surface topography was scanned using the non-contact mode with an attractive mode scan.

Supplementary Fig.~\ref{Sfig:indentation} illustrates a typical measured indentation curve. We observe that the initial snap down of the tips during the approaching phase is similar to the drumhead-suspending membrane~\cite{Lee2008Measurement,Ruiz-Vargas2011GrHole}. Furthermore, we observe an additional upward pulling of the membrane during tip retraction, which was not observed in previous suspended membrane studies. This phenomenon arises due to the freestanding \ch{MoS2} membrane adhering to the tip during indentation. The energy required to pull the water upwards is lower than the energy needed to delaminate the membrane from the tip during the extraction process. Consequently, for the analysis of the force-indentation behavior, only the curve obtained during the indentation phase was utilized.

The indentation measurements were conducted on two different grain sizes ($D = 0.1~ \mu m$ and $1.2~ \mu m$) in $10 \times 10$ grids (Fig.~\ref{fig:fig4}d,e and Supplementary Fig.~\ref{Sfig:indentation} (right) for complete results).

\subsection{Scaling analysis}

To explain morphological changes shared between the experimental and model wrinkles in Fig.~\ref{fig:fig2}, 
we propose an argument using the theory of thin sheets~\cite{Sharon2002Wavy,Klein2007Shaping,Tobasco2021Curvature}, specifically tension (or curvature) induced wrinkling~\cite{Cerda2003Geometry,Davidovitch2011Prototypical,Yifan2022Mechanics}. 
To review this mechanism, consider the simpler case of an isolated thin sheet of length $L$ and width $W$ that is subjected to an external uniaxial tension $T$ along its length and transverse compression $\delta W$ along its width, see Supplementary Fig.~\ref{fig:wrinkleschem}a. 
The sheet will wrinkle with typical amplitude $A$, width $\lambda$, and length $\ell$. 
The wrinkles relieve nearly all compressive strain, implying that $A$ and $\lambda$ obey the arc-length conservation condition:
\begin{align}
    \frac{A}{\lambda} \propto& \qty( \frac{\delta W}{ W})^{1/2} \label{eq:tiw1}
\end{align}
Moreover, in the wrinkled configuration, the bending energy density of the short wavelength ripples scales as $E_\text{bend} \propto B \qty(A/\lambda^2)^2$, and the stretching energy density due to vertical deflection scales as  $E_\text{stretch} \propto T (A/L)^2$. See Supplementary Fig.~\ref{fig:wrinkleschem}b-c for an illustration.
Minimization of these energies then implies:
\begin{align}
 \lambda \propto \qty( \frac{B}{T} )^{1/4} L^{1/2} \label{eq:tiw2}
\end{align}
Finally, the length $\ell$ of the wrinkles is proportional to the length of the tensile sheet:
\begin{align}
\ell \propto L \label{eq:tiw3}
\end{align}
Equations~(\ref{eq:tiw1}-\ref{eq:tiw3}) are the scaling relations associated with tension induced wrinkling~\cite{Cerda2003Geometry, Davidovitch2011Prototypical}. For a detailed review, we refer to Ref.~\cite{Yifan2022Mechanics}.  

We apply Eqs.~(\ref{eq:tiw1}-\ref{eq:tiw2})  to a large, unconfined membrane with disordered strain. We assume the strain is characterized by a typical variation $\Delta \varepsilon$ that occurs over a length scale $\xi$, referred to as the correlation length. 
To minimize the elastic energy, consider releasing the membrane in two steps: first release the membrane from a substrate by allowing the atoms to move in-plane, but not in the third dimension. This will reduce the initial stress to some partially relaxed stress field $\sigma_{ij}^* (\vb x)$. At every point in space the stress tensor $\sigma_{ij}^*(\vb x)$  can be decomposed into orthogonal principal axes. When one principal axis is under compression and the other is under tension, the membrane will wrinkle to alleviate the compression. In this situation, we can apply Eqs.~(\ref{eq:tiw1}-\ref{eq:tiw3}) to determine the characteristic amplitude $A$, length $\ell$, and width $\lambda$. 
To do so, we infer that (i)
the effective length $L$ of any given tensile unit is proportional to the correlation length $\xi$; (ii) the typical compression $\delta W/W$ is proportional to $\Delta \epsilon$; and (iii)
the typical tension $T$ is proportional to the Young's modulus times the strain variation $Y \Delta \varepsilon$. Substituting these identifications into Eqs.~(\ref{eq:tiw1}-\ref{eq:tiw3}) yields
\begin{align}
 \frac{A}{\lambda}& \propto \qty(\Delta \varepsilon)^{1/2} \label{eq:deriv1} \\
 \lambda 
 &\propto (\Delta \varepsilon)^{-1/4} t^{1/2} \xi^{1/2}  \label{eq:deriv2} \\
 \ell &\propto \xi \label{eq:deriv3}
\end{align}
where $t \propto \sqrt{B/Y}$ is the elastic thickness. Equations~(\ref{eq:deriv1}-\ref{eq:deriv3}) can be re-arranged to obtain Eqs.~(\ref{eq:scale1}-\ref{eq:scale3}) in the main text.

\clearpage

\setcounter{page}{1}

\begin{center}
\section*{Supplementary information for \\ 
Universal wrinkling of freestanding atomically thin films}    
\end{center}

\subsection*{Contents}
\begin{enumerate}[label=S\arabic*.]
\item Continuum model with disordered strain
  \begin{enumerate}[label*=\Alph*.]
    \item Continuum formulation
    \item Numerical discretization
  \end{enumerate}
\item Theoretical analysis of indentation experiments
\begin{enumerate}[label*=\Alph*.]
    \item Governing equations
    \item Indentation profile and qualitative expectations
    \item Derivation of mechanical resolution
    \item Estimation of effective moduli
  \end{enumerate}
\item The water container used for floating \ch{MoS2} on water.
\item Time-dependent displacements of freely floating 2D solids on water surface.
\item Comparison of AFM on water with different non-contact modes.
\item Height and phase images of \ch{MoS2} as grown on \ch{SiO2}-Si and on water.
\item Grain-size dependent surface topographies of \ch{MoS2} floating on water.
\item Formation of grain boundaries and strain distribution.
\item Force-indentation curve measurements.
\item Tension-induced wrinkling schematic.
\end{enumerate}
\clearpage

\setcounter{figure}{0}
\renewcommand{\thefigure}{S\arabic{figure}}
\renewcommand\figurename{Fig.}

\setcounter{table}{0}
\renewcommand{\thetable}{S\arabic{table}}

\renewcommand{\theequation}{S\arabic{equation}}
\setcounter{equation}{0}
\renewcommand{\thesection}{S\arabic{section}}
\renewcommand{\thesubsection}{S\arabic{subsection}.}
\setcounter{subsection}{0}
\renewcommand{\thesubsubsection}{S\arabic{subsection}.\Alph{subsubsection}}

\subsection{Continuum model with disordered strain} 

Here we detail the elastic model in Fig.~\ref{fig:fig2}c-d. The model consists of a free-floating thin elastic sheet with a finite bending modulus $B$ and Young's modulus $Y$;  no external forces (e.g. surface tension or gravity) are included.
For clarity, we first present the model in a continuum formulation, and then we present a discretized version amenable to numerical relaxation. 
 Our implementation of the model retains full geometric nonlinearity; it does not rely on a shallow deformation approximation or other simplifications prior to numerical discretization.

\subsubsection{Continuum formulation}

In the continuum, we specify the geometry of the thin sheet using a coordinate system $s \in \RR^2$ with an embedding  $\vec r : \RR^2 \to \RR^3$. We take the total energy to be the sum of bending and stretching energy:
\begin{align}
    E = E_\text{stretch} + E_\text{bend} 
\end{align}
The bending energy is given by:
\begin{align}
    E_\text{bend} = B \int  \Tr(N^2) \sqrt{ \det g}\, \dd^2 s
\end{align}
where $B$ is the bending modulus, and $g$ and $N$ are the first and second fundamental forms, respectively. The fundamental forms are expressed in terms of the embedding $\vec r (s)$ as
\begin{align}
g_{ab} = \partial_a \vec r \cdot \partial_b \vec r \quad \text{and} \quad  N_{a b } =  \partial_a \vec r \cdot \partial_b  \hat n 
\end{align}
where 
\begin{align}
 \hat n = \frac{\vec n}{\abs{\vec n}}  \quad \text {and} \quad \vec n = \frac12 \epsilon_{ab} \partial_a \vec r \times \partial_b \vec r
\end{align}
Here, $\epsilon_{ab}$ is the Levi-Civita symbol, $\partial_a$ denotes $\pdv{s^a}$, and repeated indices are summed over.  The stretching energy we use is given by. 
\begin{align}
    E_\text{stretch} = \frac18 \int \left[ \lambda_1 \left(\Tr \Delta g \right)^2 +\lambda_2 \Tr \left( \Delta g \right)^2  \right] \sqrt{ \det g} \, \dd^2s 
\end{align}
where $\lambda_1$ and $\lambda_2$ are Lam\'e coefficients. In terms of Lam\'e coefficients, the Young's modulus $Y$ is given by 
\begin{align}
    Y = \frac{  4\lambda_2 (\lambda_1 + \lambda_2) }{\lambda_1+2 \lambda_2} 
\end{align}
The quantity $\Delta g$ is the metric deviation $\Delta g = g - g_0$, where $g_0$ is a target metric. 

The target metric is specified in a distinguished coordinate system as follows. For each point $s$, draw two values $\nu_\parallel (s) $ and $\nu_\perp (s)$ independently from the uniform distribution $[-1,1]$, and draw an angle $\theta(s)$ from the uniform distribution over $[0,\pi) $. Define the matrix:
\begin{align}
    \eta(s) = \mqty[ \nu_\parallel \cos^2 \theta+\nu_\perp \sin^2 \theta & (\nu_\perp - \nu_\parallel) \cos \theta \sin \theta \\ 
    (\nu_\perp -\nu_\parallel) \cos \theta \sin \theta & \nu_\parallel \sin^2 \theta + \nu_\perp \cos^2 \theta 
    ]
\end{align}
which is the matrix with eigenvalues $\nu_\parallel$ and $\nu_\perp$ and eigenvectors $[\cos \theta, \sin \theta]^T$ and $[-\sin \theta, \cos \theta]^T$, respectively. 

Next, we introduce a strain scale $\Delta \varepsilon$ and a correlation length $\xi$ and define 
\begin{align}
    g_0(s)  = 1+  \frac{\xi \Delta \varepsilon}{Z (s) } \int \eta(s') \exp[-\frac{ d(s,s')^2 }{2 \xi^2}   ] \dd^2 s' \label{eq:metric}
\end{align}
where $Z(s) = \int  \exp[-d(s,s')^2 / 2 \xi^2 ] \dd s' $ is a normalization constant, and 
\begin{align}
d(s,s') =\sqrt{ (s_1-s'_1)^2 + (s_2-s_2')^2 } 
\end{align}
imposes a Euclidean norm on the the distinguished coordinate system. (If the sheet has infinite size, then $Z (s) 
 = 2 \pi \xi^2 $). The model is relaxed in order to find embeddings that minimize the total energy $E$ with free boundary conditions. Our interest is in the regime of an asymptotically thin sheet, i.e. the elastic thickness $t \propto \sqrt{B/Y}$ is much smaller than any other length scale, such as the system size or correlation length. In order to empirically find the minimizers, we use a discrete approximation and numerical relaxation described in the next section.  

\subsubsection{Numerical discretization}
 
In the discrete setting, the thin sheet is represented by a collection of vertices $\VV$, each of which is assigned a reference point $\vec x_i \in \RR^2$ in the flat Cartesian plane, and a target point $\vec r_i \in \RR^3 $ in three dimensions. The reference points $\vec x_i$ form a regular triangular lattice of lattice spacing $a$. 
The vertices are triangulated to form a set $\NN$ of counterclockwise oriented triangular simplices $\alpha = (\alpha_1, \alpha_2, \alpha_3) \in \VV^3$. 
The total energy is given by:
\begin{align}
    E = E_\text{stretch}+ E_\text{bend}
\end{align}
The bending energy is given by a standard normal deflection term~\cite{Seung1988Defects}:
\begin{align}
    E_\text{bend} = \frac{\kappa}{2}  \sum_{\alpha \in \NN }\sum_{\beta \in N(\alpha) } \qty(1- \hat n_\alpha \cdot \hat n_\beta)
\end{align}
where $\hat n_\alpha \in \RR^3$ is the normal vector to simplex $\alpha$ and $N(\alpha)$ denotes the set of all simplices that share an edge with simplex $\beta$. The normals can be computed via the shoelace formula:
\begin{align}
    \hat n_\alpha = \frac{\vec n_\alpha }{\abs{\vec n_\alpha} }  \quad \text{with} \quad \vec n_\alpha = \frac12 \sum_{i=1}^3 \vec r_{\alpha_i} \times \vec r_{\alpha_{i+1}}
\end{align}
where the indices on the simplex $\alpha$ are interpreted periodically. In terms of $\kappa$, the macroscopic bending stiffness is $B=\sqrt{3} \kappa/2$~\cite{Seung1988Defects}.

The stretching energy is given by:
\begin{align}
    E_\text{stretch} = \frac1{8} \sum_{\alpha \in \NN} A^\alpha \left[ \lambda_1 \left(  \Tr \Delta g^\alpha \right)^2  + \lambda_2 \Tr (\Delta g^\alpha)^2 \right]
\end{align}
Here, $A^\alpha$ is the area of simplex $\alpha$, which we approximate by its undeformed value $A^\alpha \approx \frac{\sqrt{3}}{2} a^2$. The quantity $\Delta g^\alpha =g^\alpha - g_0^\alpha $ is the metric deviation (i.e., twice the strain tensor). Here, $g^\alpha$ is the discrete metric computed for each simplex. In the coordinate system  $\vec x_i$, it is given by:
\begin{align}
    g^\alpha = X_\alpha^{-T} \cdot  M_\alpha \cdot X^{-1}_\alpha
\end{align}
The matrix $X$ is defined as 
\begin{align}
    X_\alpha = \mqty[ \vec x^{12}_\alpha  & \vec x^{23}_\alpha ]
\end{align}
where $\vec x^{12}_\alpha = \vec x_{\alpha_1} - \vec x_{\alpha_2}$ and 
$\vec x^{23}_\alpha = \vec x_{\alpha_2} - \vec x_{\alpha_3}$. The matrix $M_\alpha$ is defined as 
\begin{align}
M_\alpha = \mqty[ \vec r^{12}_\alpha \cdot \vec r^{12}_\alpha & \vec r^{23}_\alpha \cdot \vec r^{12}_\alpha \\
\vec r^{23}_\alpha \cdot \vec r^{12}_\alpha & \vec r^{23}_\alpha \cdot \vec r^{23}_\alpha]
\end{align}
where $\vec r_\alpha^{12} = \vec r_{\alpha_1} -\vec r_{\alpha_2}$ and $\vec r_\alpha^{23} = \vec r_{\alpha_2} - \vec r_{\alpha_3}$. 

The target metric $g_0^\alpha $ is constructed in a similar manner as its continuum counterpart in Eq.~(\ref{eq:metric}). For each simplex, we draw two values $\nu_\parallel^\alpha$ and $\nu_\perp^\alpha$ from the uniform distribution over $[-1,1]$, and an angle $\theta^\alpha $ from the uniform distribution over $[0, \pi)$. We define the matrix $\eta^\alpha$ as:
\begin{align}
    \eta^\alpha = \mqty[ \nu_\parallel^\alpha \cos^2 \theta^\alpha +\nu_\perp^\alpha \sin^2 \theta^\alpha & (\nu_\perp^\alpha - \nu_\parallel^\alpha) \cos \theta^\alpha \sin \theta^\alpha \\ 
    (\nu_\perp^\alpha -\nu_\parallel^\alpha) \cos \theta^\alpha \sin \theta^\alpha & \nu_\parallel^\alpha \sin^2 \theta^\alpha + \nu_\perp^\alpha \cos^2 \theta^\alpha 
    ]
\end{align}
which is the matrix with eigenvalues $\nu_\parallel^\alpha$ and $\nu_\perp^\alpha$ and orthogonal eigenvectors $[\cos \theta^\alpha, \sin \theta^\alpha]^T$ and $[-\sin \theta^\alpha, \cos \theta^\alpha]^T$, respectively. To construct the target metric, we specify a correlation length $\xi$ and a strain magnitude $\Delta \varepsilon$, and  define
\begin{align}
    g^\alpha_0 = 1+ \frac{ \xi \Delta \varepsilon }{Z_\alpha} \sum_{\beta \in \NN} \eta^{\beta} \exp(- \frac{d_{\alpha\beta}^2 }{2 \xi^2   }  )
\end{align}
where $d_{\alpha \beta} = \left| \frac13 \sum_{i=1}^3 ( \vec x_{\alpha_i} - \vec x_{\beta_i}) \right|$ is the distance between the centers of simplices $\alpha$ and $\beta$, and $Z_\alpha = \sum_{\beta \in \NN} \exp(-  d_{\alpha \beta}^2 / 2 \xi^2 )  $ is a normalization constant. The energy is minimized using the Fast Inertial Relaxation Engine~~\cite{Bitzek2006Structural,Hanakata2021Thermal}.

\subsection{Theoretical analysis of indentation experiments}
\begin{table}[t!]
\centering
\begin{tabular}{|c||l|l|}
    \hline 
    {\bf Symbol} & {\bf Physical meaning} & {\bf Value}\\ 
    \hline 
    $\gamma$  & Air-water surface tension & $\approx 7.2 \times 10^{-2} kg/s^2$ \\
   \hline 
   $B_0$ & Bending modulus of single-crystal \ch{MoS2}  & $\approx 1.7 \times 10^{-18} kg \, m^2/s^2$ \\
   \hline
    $Y_0$ & Young's modulus of single-crystal \ch{MoS2} & $\approx 1.7 \times 10^{2} kg /s^2$ \\
    \hline 
    $\rho g$ & Gravitational force of water & $\approx 10^4 kg/m^2s^2$ \\
    \hline
    $R$ & Radius of floating membrane &  $\approx 2\times 10^{-4} m$ \\
    \hline 
    $r_\text{tip}$ & AFM tip radius & $\approx 5 \times 10^{-9} m$ \\
    \hline 
    $\nu$ & Poisson's ratio of \ch{MoS2} & $\approx 0.25$ \\
    \hline 
\end{tabular}
\caption{{\bf Physical parameters entering indentation analysis.}}
\label{tab:params}
\end{table}

\subsubsection{Governing equations}
We analyze the AFM indentation experiments through the lens of an effective medium theory in which the wrinkled, free-floating membrane is approximated as a uniform membrane with renormalized effective bending stiffness $B_\text{eff}$ and stretching stiffness $Y_\text{eff}$. For small angle deformations, the membrane is described by the F\"oppl-von K\'arm\'an equations:
\begin{align}
    B_\text{eff} \nabla^4 h + \epsilon_{ij} \epsilon_{kl} \partial_i \partial_k \chi \partial_j \partial_l h + \rho g h =& 0 \label{eq:fvk1} \\ 
    \nabla^4 \chi  + \frac{Y_\text{eff}}{2} \epsilon_{ij} \epsilon_{kl} \partial_i \partial_k h \partial_j \partial_l h =& 0 \label{eq:fvk2} 
\end{align}
where 
$h$ is the height of the membrane, $\rho g h$ is the force of gravity, and $\chi$ is the Airy stress function. 

We consider a circular membrane of radius $R$ on the surface of water subject to an indentation of depth $\delta$ by a circular indenter of radius $r_\text{tip}$. Outside the membrane, there is a narrow ($\approx 1 \mu m$) channel of exposed water surrounded by another segment of \ch{MoS2} that is anchored to a solid lid. We will assume that the outer segment of \ch{MoS2} pins the water level to $0$. Anticipating azimuthal symmetry of the indented membrane, Eqs.~(\ref{eq:fvk1}-\ref{eq:fvk2}) become:
\begin{align}
    \frac{B_\text{eff}}{r}\dv{r} \qty{ r \dv{r} \qty[ \frac1r \dv{r} \qty(r \dv{h}{r}) ] }- \frac1r \dv{r} \qty(\dv{\chi}{r} \dv{h}{r} )  + \rho g h = 0  \label{eq:radial1} \\
     \dv{r} \qty{ r \dv{r}\qty[ \frac1r \dv{r} \qty( r  \dv{\chi}{r})] }+ \frac{ Y_\text{eff}}{2} \dv{r} \qty( \dv{h}{r} )^2 =0 \label{eq:radial2}
\end{align}
Subject to azimuthal symmetry, the stress tensor is related to the Airy stress function via:
\begin{align}
    \sigma_{rr}  = \frac{1}{r} \dv{\chi}{r}   \quad \text{and} \quad \sigma_{\theta \theta} = \dv[2]{\chi}{r}
\end{align}
The assumption of azimuthal symmetry is self-consistent so long as $\sigma_{\theta \theta} >0$ for all $r$. If $\sigma_{\theta \theta} < 0$, the membrane undergoes a spontaneous symmetry breaking via an azimuthal buckling instability~\cite{Box2017waterIndent,Vella2018WaterIndent}. We will show below that our indentations remain well below this buckling instability. Boundary conditions and the self-consistency of the effective medium picture are discussed below. Tables of relevant parameters and length scale are provided in Tables~\ref{tab:params}-\ref{tab:lengths}, respectively.

\begin{table}[t!]
\centering
\begin{tabular}{|c||c|l|l l |}
    \hline 
    {\bf Symbol} &  {\bf Expression }& {\bf Physical meaning} & {\bf Value ($m$) } &\\ 
    \hline 
    $t$ &  $\sqrt{12(1- \nu^2 )\frac{B_\text{eff}}{Y_\text{eff}}}$  & Elastic thickness & $\approx 3 \times 10^{-10} $ & for $B_\text{eff} = B_0$ \\ 
     & & & & and $Y_\text{eff} = Y_0$ \\
    \hline
    $r_\text{tip}$ & --- &  AFM tip radius & $\approx 5 \times 10^{-9} $ & \\
   \hline 
    $\ell_\text{EC}$ & $\sqrt{\frac{B_\text{eff}}{\gamma}}$ & Elasto-capillary length & $\approx 5 \times 10^{-9} $ & for $B_\text{eff}= B_0$  \\
    \hline 
    $\ell_g $& $\qty( \frac{B_\text{eff}}{\rho g})^{1/4}$ & Elasto-gravitational length & $\approx 3.6 \times 10^{-6} $ & for $B_\text{eff} = B_0$  \\
    \hline 
    $\delta_\text{max}$ & --- & Maximum indentation depth & $\approx 5 \times 10^{-7} $ & \\
    \hline 
    $R$ & --- &  Membrane radius & $\approx 2 \times 10^{-4}$ & \\
    \hline
    $\ell_\text{C}$ & $\sqrt{\frac{\gamma}{\rho g}}$  & Capillary length & $ \approx 2.3 \times 10^{-3} $ & \\
    \hline 
\end{tabular}
\caption{{\bf Length scales entering indentation analysis.} For lengths that depend on effective moduli $B_\text{eff}$ and $Y_\text{eff}$, reference values are provided for single-crystal \ch{MoS2} with moduli $B_0$ and $Y_0$. For a wrinkled sheet, one generally expects $B_\text{eff} > B_0$ and $Y_\text{eff} < Y_0$. }
\label{tab:lengths}
\end{table}

\subsubsection{Indentation profile and qualitative expectations}

To understand the dominant contributions to Eq.~(\ref{eq:radial1}-\ref{eq:radial2}), we introduce the following nondimensionalization:
\begin{align}
    \xi =& \frac{r}{R} &
    \tilde \chi =&  \frac{\chi}{\gamma R^2} &
    \tilde h =& \frac{h}{\delta}
\end{align}
in terms of which Eqs.~(\ref{eq:radial1}-\ref{eq:radial2}) become
\begin{align}
      \frac{\ell_\text{EC}^2}{R^2}\frac{1  }{\xi}\dv{\xi} \qty{ \xi \dv{\xi} \qty[ \frac1\xi \dv{\xi} \qty(\xi \dv{ \tilde h}{\xi}) ] }- \frac1\xi \dv{\xi} \qty(\dv{\tilde \chi}{\xi} \dv{\tilde h}{\xi} )  + \frac{R^2 }{\ell_\text{C}^2 } \tilde h =& 0  \label{eq:ndim1} \\
     \dv{\xi} \qty{ \xi \dv{\xi}\qty[ \frac1\xi \dv{\xi} \qty( \xi  \dv{ \tilde \chi}{\xi})] }+ \frac{Y}{2 \gamma} \left( \frac{\delta}{R} \right)^2   \dv{\xi} \qty( \dv{ \tilde h}{\xi} )^2 =& 0 \label{eq:ndim2}
\end{align}
where $\ell_\text{C}$ and $\ell_\text{EC}$ are, respectively, the capillary and elasto-capilary lengths given in Table~\ref{tab:lengths}.
Unlike macroscopic membrane floating experiments~\cite{Box2017waterIndent}, our membranes are much smaller than the capillary length, so the buoyancy contribution, $(R/\ell_\text{C})^2 \approx 10^{-2}$, is negligible. Next, notice that $(\ell_\text{EC}/R)^2$ implicitly depends on $B_\text{eff}$. For $B_\text{eff} = B_0$, one has  $(\ell_\text{EC}/R)^2 \approx 10^{-10}$. Therefore, even if $B_\text{eff}$ is several orders of magnitude larger than $B_0$, the bending term is negligible outside of a small effective internal radius $r_{in} \approx r_\text{tip}+\ell_\text{EC}$. Thus, Eqs.~(\ref{eq:ndim1}-\ref{eq:ndim2}) reduce to: 
\begin{align}
        \dv{\xi} \qty(\dv{\tilde \chi}{\xi} \dv{\tilde h}{\xi} )  =& 0 \label{eq:reduced1} \\
      \xi \dv{\xi}\qty[ \frac1\xi \dv{\xi} \qty( \xi  \dv{ \tilde \chi}{\xi})] + \Lambda   \qty( \dv{ \tilde h}{\xi} )^2 = & 0 \label{eq:reduced2}
\end{align}
where we have set the constant of integration between Eq.~(\ref{eq:ndim2}) an Eq.~(\ref{eq:reduced2}) to zero as is commonly assumed~\cite{Komaragiri2005Mechanical}. 
Notice that Eqs.~(\ref{eq:reduced1}-\ref{eq:reduced2}) depend only on a single dimensionless parameter
\begin{align}
    \Lambda= \frac{Y_\text{eff}}{2 \gamma} \qty( \frac{\delta}{R})^2
\end{align}
which summarizes the indentation depth relative to the membrane diameter and the elastic stiffness relative to the pre-tension. 
To compare with other forms of indentation experiments, we solve Eqs.~(\ref{eq:ndim1}-\ref{eq:ndim2}) subject to two distinct sets of boundary conditions. First, as appropriate for floating membranes, we require:  
\begin{align}
\tilde h (\tilde r_{in}) =& -1 \label{eq:s1c1}\\
\tilde h(1) =& 0 \label{eq:s1c2} \\
\left( \dv[2]{\tilde \chi}{\xi} - \frac{\nu}{\xi} \dv{\tilde \chi}{\xi}  \right)_{\xi = \tilde r_{in}} =& 1 - \nu \label{eq:s1c3} \\
\eval{\dv{\tilde \chi}{\xi}}_{\xi= 1} =& 1 \label{eq:s1c4}  
\end{align}
Second, as appropriate for drumhead-like, suspended membranes, all the boundary conditions are identical except that Eq.~(\ref{eq:s1c4}) is replaced with 
\begin{align}
\left( \dv[2]{\tilde \chi}{\xi} - \frac{\nu}{\xi} \dv{\tilde \chi}{\xi}  \right)_{\xi =1} =& 1 - \nu  \label{eq:s2c4}  
\end{align}
Equations~(\ref{eq:s1c1}-\ref{eq:s1c2}) set the membrane height to $-\delta$ and $0$ at $r_\text{in}$ and $R$, respectively. Equation~(\ref{eq:s1c3}) is a no radial slip condition at $r= r_\text{in}$. Equation~(\ref{eq:s1c4}) imposes a constant stress boundary condition outer edge: $\sigma_{rr}(R)=\gamma$. By contrast, Eq.~(\ref{eq:s2c4}) is a no-slip condition at the outer edge. Equation~(\ref{eq:s1c4}) is appropriate for membranes floating on water, and Eq.~(\ref{eq:s2c4}) is appropriate for drumhead-like suspended membranes, where the outer boundary is pinned via van der Waals adhesion to a solid substrate. (For drumhead-like suspended membranes, $\gamma$, used to nondimensionize $\chi$, represents the typical pre-tension arising during the suspension process, not the surface tension of water.)    

\begin{figure}
    \centering
    \includegraphics[width=\textwidth]{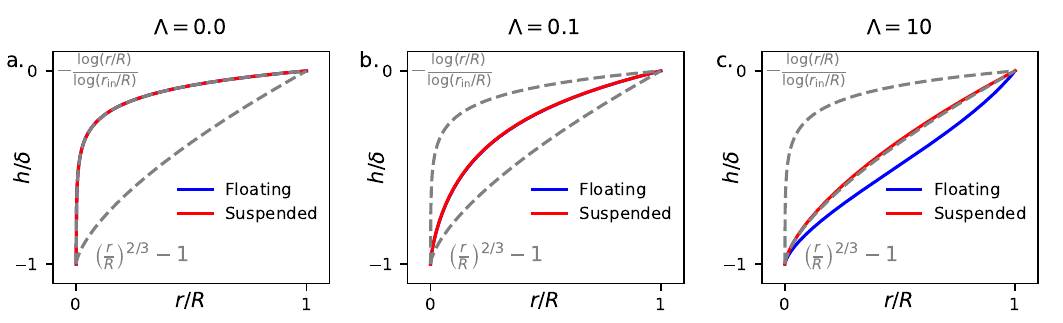}
    \caption{{\bf Boundary conditions and indentation regimes.} The spatial profiles $h(r)$ are shown for increasing values of $\Lambda = \frac{Y}{2\gamma}\qty(\frac{\delta}{R})^2$. 
    The red curves correspond to a pinned boundary condition, as is appropriate for drumhead-like suspended membrane experiments. The blue curves correspond to constant stress boundary conditions, applicable for floating membranes. Analytical asymptotic solutions are plotted as dashed grey lines. For these plots $\nu =1/3$ and $\tilde r_\text{in} = 10^{-4}$.  }
    \label{fig:regimes}
\end{figure}

In Fig.~\ref{fig:regimes}, we plot the the height profile $\tilde h (\xi)$ for three values of $\Lambda$, obtain by numerical integration of Eqs.~(\ref{eq:ndim1}-\ref{eq:ndim2}).  Notice that for $\Lambda \ll 1$, the suspended and floating profiles are nearly identical, while for larger $\Lambda \gg 1$ the profiles are distinct. 
The dashed lines in Fig.~\ref{fig:regimes} represent analytical solutions are available in two limiting cases. 
When $\Lambda \gg 1$ (and $\nu =1/3$), one obtains~\cite{Komaragiri2005Mechanical}:
\begin{align}
\tilde h \approx \tilde \xi ^{2/3} -1  \quad \text{and} \quad 
\tilde \chi \approx  \frac{2}{3} \Lambda \xi^{4/3}
\end{align}
for the suspended boundary conditions. This regime is relevant for deep ($\sim 1 \mu m$) indentation of drumhead-like suspended atomically thin films, 
for which  $ R \lesssim 2 \mu m$ and $\Lambda \gtrsim 50$~\cite{Lee2008Measurement}. 
In this case, the measured force [$F = \eval{ 2\pi r \sigma_{rr} \dv{h}{r}}_{r_\text{in}}$] exhibits a cubic relationship with indentation depth: $F \approx  ( 8 \pi Y_\text{eff}/9 R^2  ) \delta^3 $. Second, when $\Lambda =0$, one obtains 
\begin{align}
    \tilde h =& -\frac{\log(\xi)}{\log(\tilde r_{in})} \quad \text{and} \quad 
    \tilde \chi = \frac{\xi^2}{2} \label{eq:sol2}
\end{align}
which is a valid approximation for extremely shallow ($\delta \lesssim t$) indentations of both floating and suspended membranes. This limit also describes the indentation of the surface of water when no membrane is present (namely, $Y_\text{eff} = B_\text{eff} =0$ and so $\Lambda =0$ and $r_\text{in} = r_\text{tip}$).  Equation~(\ref{eq:sol2}) predicts a linear force-displacement relations $F = 2 \pi \gamma \delta / \log (R/r_\text{in})$. Using the values from Table~\ref{tab:params}, our floating membrane experiments exist in the regime $\Lambda \lesssim 0.01$.  
This implies that our observed super-hookean force response (see Fig.~\ref{fig:fig4}a) can be decomposed into two contributions: (i) a dominant linear response set by the surface tension $\gamma$ and modulated logarithmically by the ratio $R/r_\text{in}$, which depends implicitly on $r_\text{tip}$ and $B_\text{eff}$, and (ii) a superhookean correction due to the finite Young's modulus $Y_\text{eff}$ entering $\Lambda$. As detailed in \S\ref{sec:estimate}, we use the intensity of the superhookean correction to estimate $Y_\text{eff}$. 

Finally, we address the possibility of azimuthal buckling. Notice that Eq.~(\ref{eq:ndim1}) may be re-written as 
\begin{align}
\dv[2]{\tilde h}{\xi} = - \frac{ \tilde \sigma_{\theta \theta }}{\tilde \sigma_{rr}} \dv{\tilde h}{\xi} \label{eq:inflection}
\end{align}
where $\tilde \sigma = \sigma/\gamma$ is the nondimensionzlied stress. Importantly, Eq.~(\ref{eq:inflection}) states that $\sigma_{\theta \theta} =0$ only if $\tilde h$ has an inflection point. For $\Lambda \ll 1$, the regime of our floating membranes, a perturbative analysis of Eq.~(\ref{eq:sol2}) or visual inspection of Fig.~\ref{fig:regimes} reveal that no inflection point is present. Hence, the assumption of azimuthal symmetry is self-consistent in the effective medium picture.

\subsubsection{Derivation of mechanical resolution}

\begin{figure}[t!]
    \centering
    \includegraphics[width=0.8\textwidth]{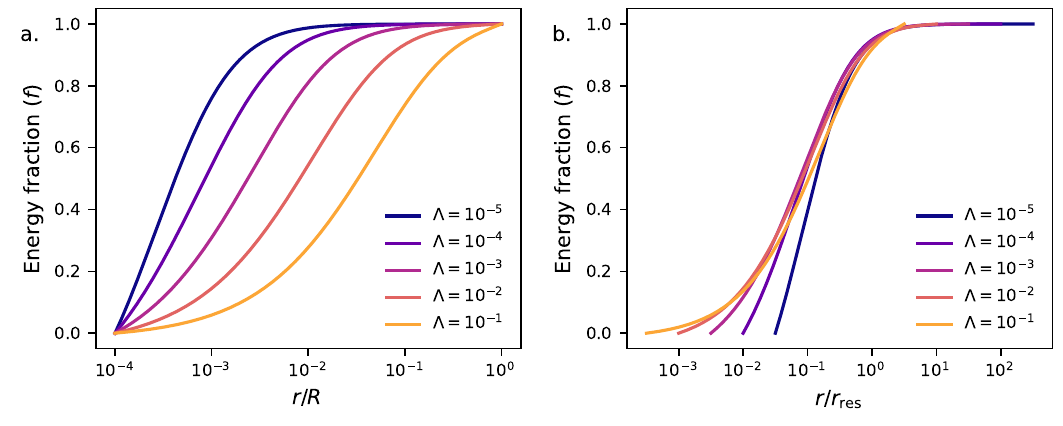}
    \caption{ {\bf Identification of mechanical resolution.}~{\bf a.} Fraction $f$ of deformation energy as a function of $\xi = r/R$ for $\Lambda \ll 1$, see Eq.~(\ref{eq:fraction}). The shoulder in each curve sets a radius, the mechanical resolution, within which the majority of the deformation energy resides.~{\bf b.} The location of the shoulder can be collapsed by rescaling $r$ by $r_\text{res} = \delta \sqrt{Y/2\gamma}$.  }
    \label{fig:mechres}
\end{figure}

Here we discuss the validity of the effective medium approximation and the notion of a mechanical resolution for stiffness imaging. Since our experiments are in the regime $\Lambda \ll 1$, it is sensible to decompose the stress tensor $\tilde \sigma = \sigma/\gamma$ into a prestress (i.e. surface tension) and an additional stress: $\tilde \sigma = 1 + \tilde \Sigma$. The deformation energy density that is independent of the prestress is given by:
\begin{align}
e(\xi) = \tilde \Sigma_{rr}^2 + \tilde \Sigma_{\theta \theta}^2 - 2 \nu \tilde \Sigma_{rr} \tilde \Sigma_{\theta \theta}
\end{align}
Then the fraction of the total deformation energy stored within a radius $\xi$ is given by:
\begin{align}
f (\xi) = \frac{1}{N}\int_{\tilde r_{in}}^{\xi} e(\xi') \xi' \dd \xi' \label{eq:fraction}
\end{align}
where $N = \int_{\tilde r_{in}}^{1} e(\xi') \xi' \dd \xi ' $ is a normalization constant. In Fig.~\ref{fig:mechres}, the function $f(\xi)$ is plotted for several values of $\Lambda$. Each curve features an elbow within which roughly 90\% of the mechanical energy is stored. We define the location of this elbow as the mechanical resolution $r_\text{res}$. We can deduce $r_\text{res}$ based on the following physical argument: In physical units, for large $\Lambda$ the location of the elbow $r_\text{res}$ should be independent of membrane radius $R$. Moreover, in dimensionless variables, $r_\text{res}/R$ should depend only on $\Lambda$. These two conditions together imply: $r_\text{res} \propto  \sqrt{\Lambda} R = \delta \sqrt{ \frac{Y}{2 \gamma} }$. A collapse in Fig.~\ref{fig:mechres} shows favorable agreement. For a polycrystalline membrane with domain size $D$, we expect the effective medium picture to provide a reasonable approximation if $r_\text{res} \gg D$, since many domains participate in the indentation.  For $r_\text{res} \ll D$, we expect point probes to result in hetergeneous profiles since only a small number of grains (and therefore wrinkles) participate in a given indentation.

\subsubsection{Estimation of effective moduli}
\label{sec:estimate}

To estimate the order of magnitude of the effective Young's modulus, we numerically integrate Eq.~(\ref{eq:radial1}-\ref{eq:radial2}). Doing so requires imposing boundary conditions on $h(r)$ and $\chi(r)$.
The boundary conditions on $h(r)$ include:
\begin{align}
    h(r_\text{tip}) =& - \delta \label{eq:cond1} \\
    \eval{\dv{h}{r}}_{r=r_\text{tip}} =& 0 \label{eq:cond2} \\
    h(R) =& 0 \label{eq:cond3} \\
    \eval{\dv{r} \qty( r \dv{h}{r})}_\text{r=R}  =& 0 \label{eq:cond4}
\end{align}
Condition~(\ref{eq:cond4}) states that the water does not exert a torque at the edge of the membrane. The boundary conditions on $\chi$ include:
\begin{align}
    \eval{\dv{\chi}{r}}_{r=R} =& R \gamma \label{eq:cond5} \\
     \qty( r\dv[2]{\chi}{r} - \nu \dv{\chi}{r} )_{r=r_\text{tip}}=& \gamma (1-\nu) \label{eq:cond6} \\
     \dv{r} \qty[ \frac1r \dv{r} \qty( r \dv{\chi}{r} ) ]_{r=r_\text{tip}} =& 0   \label{eq:cond7} \\
     \chi(R) =& 0 \label{eq:cond8}
\end{align}
Condition~(\ref{eq:cond5}) states that the the outer boundary of the membrane is in mechanical equilibrium with the surface tension of the water. Condition~(\ref{eq:cond6}) states that there is no radial slip at the edge of the indenter. Condition~(\ref{eq:cond7}) is a common assumption that allows for a simplified first integral of Eq.~(\ref{eq:radial1})~\cite{Komaragiri2005Mechanical}.  Condition~(\ref{eq:cond8}) is an arbitrary gauge fixing of $\chi$, which may be omitted in practice since we need only solve for the derivatives of $\chi$. 

Given a solution for $h$, the reaction force on the indenter is given by
\begin{align}
    F = 2 \pi B_\text{eff} \qty{  r \dv{r} \qty[ \frac1{r} \dv{r} \qty(r \dv{h}{r})]}_{r = r_\text{tip}}
\end{align}
The black, yellow, and blue dashed theoretical curves in Fig.~\ref{fig:fig4}a are produced by numerically solving Eqs.~(\ref{eq:fvk1}-\ref{eq:fvk2}) with $B_\text{eff} = B_0$ and $Y_\text{eff} = Y_0$, $Y_0/10$, and $Y_0/100$, respectively. The dashed grey line (corresponding to no membrane) is given by the force-displacement relationship $F = 2 \pi \gamma \delta /\log(r_\text{tip}/R)$ for the open surface of water~\cite{Box2017waterIndent}.

\clearpage

\begin{figure*}[t!]
\centering 
\renewcommand\thefigure{S\arabic{figure}}    
\includegraphics[width=\textwidth]{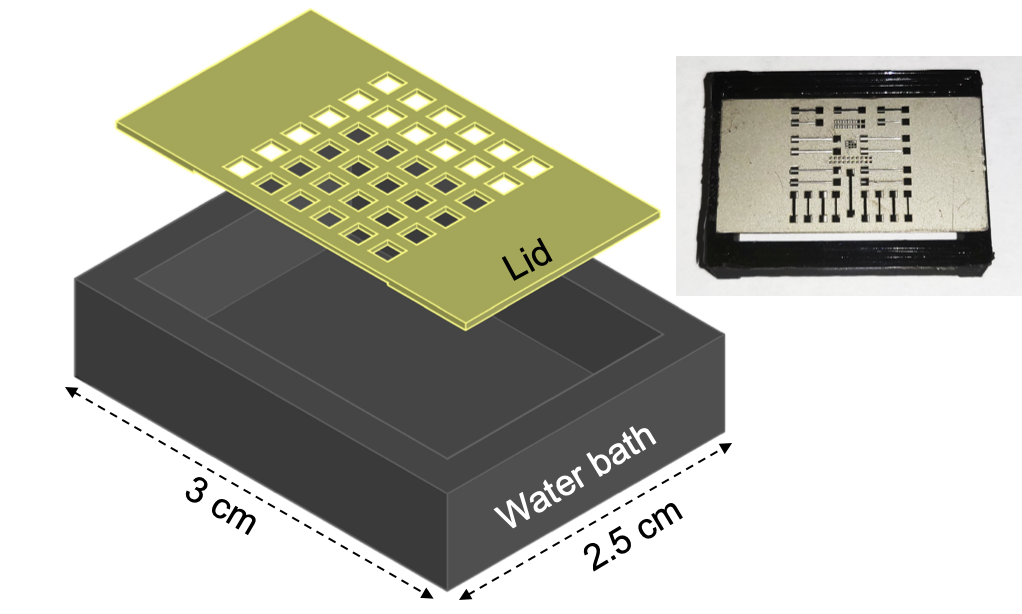}
 \caption{{\bf The water container used for floating \ch{MoS2} on water.} The design (left) and actual picture (right) of the water container with a lid. The container is fabricated using 3D printing technology with Polyactic Acid (PLA), while the lid is fabricated by laser-cutting nickel-brass thin sheets with a thickness of less than 1 mm. The holes in the lid have sizes ranging from 200 um to 2 mm.
  }
  \label{Sfig:container}
\end{figure*}

\clearpage 

\begin{figure*}[t!]
\centering 
\renewcommand\thefigure{S\arabic{figure}}    

\includegraphics[trim={7cm 15cm 4cm 0},clip,width=\textwidth]{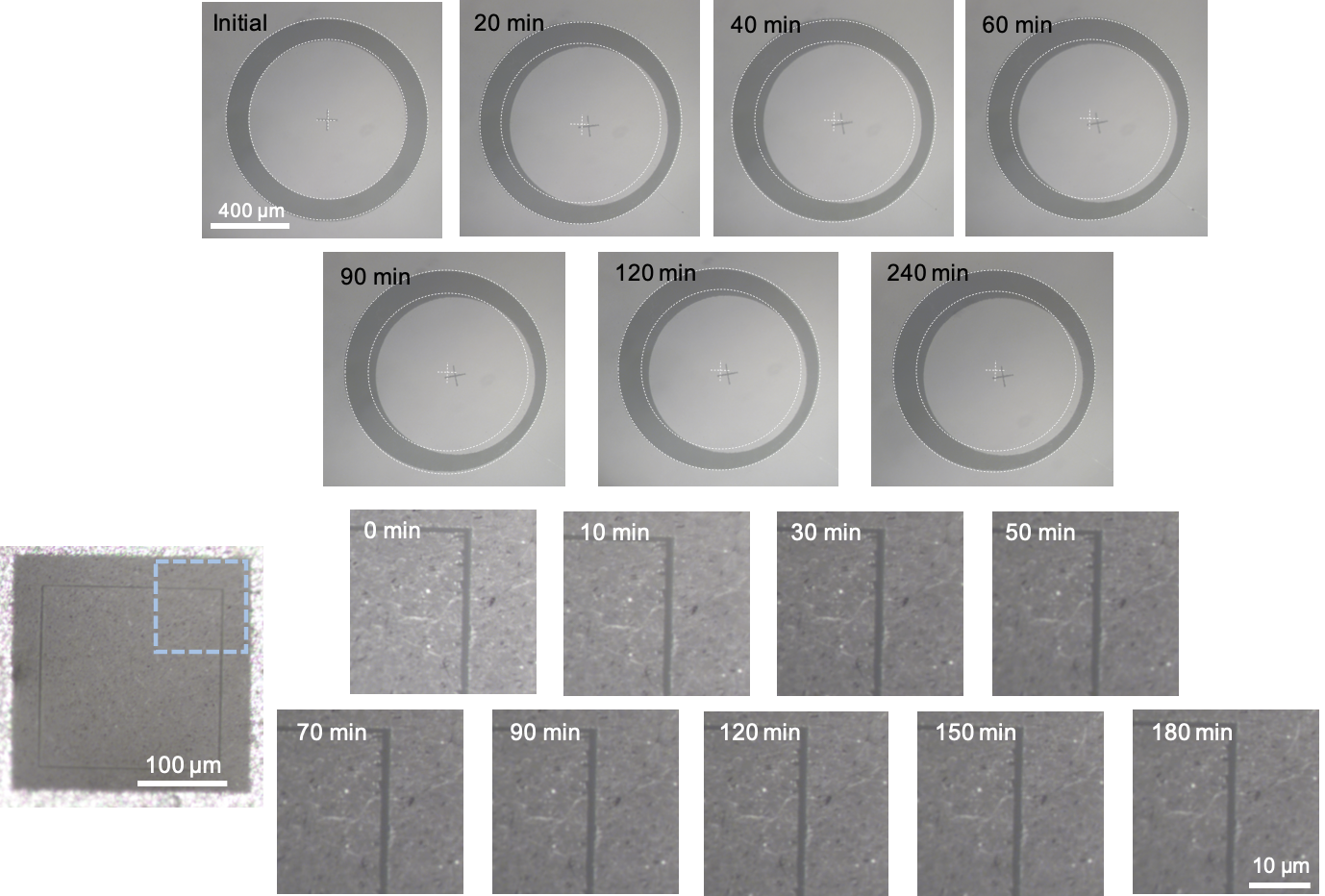}
 \caption{{\bf Time-dependent displacements of freely floating 2D solids on water surface.} 
 Translocation of the floating \ch{MoS2}, size of $600 ~\mu m$, on water. 
  }
  \label{Sfig:drift}
\end{figure*}

\clearpage 

\begin{figure*}[t!]
\centering 
\renewcommand\thefigure{S\arabic{figure}}    
\includegraphics[width=\textwidth]{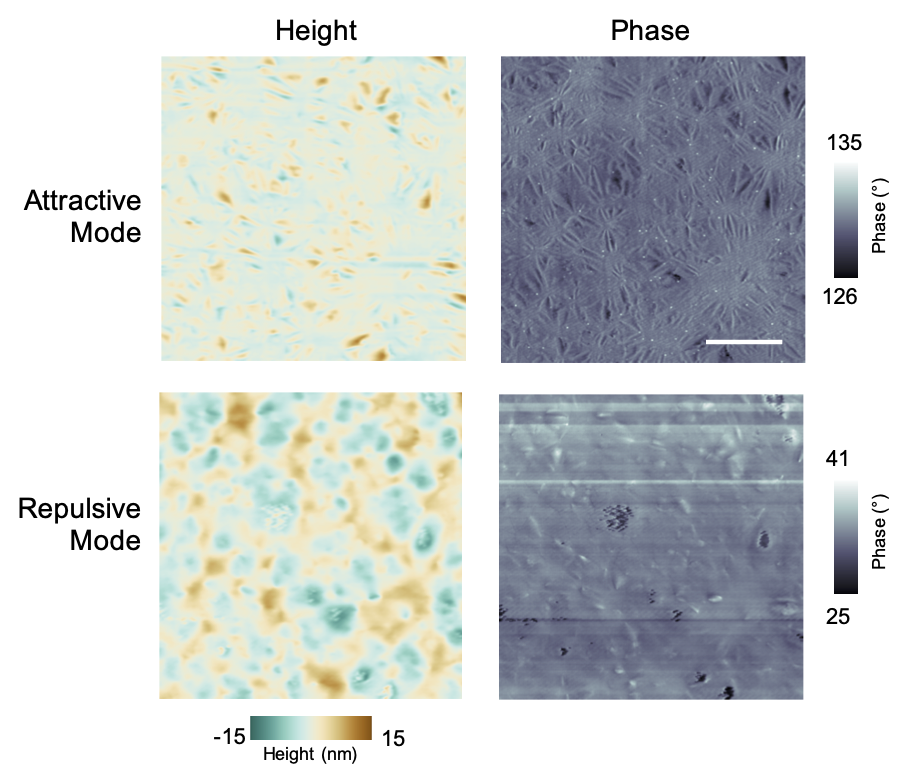}
 \caption{{\bf Comparison of AFM on water with different non-contact modes.} 
(Top) Height (left) and  phase (right) images of floating \ch{MoS2} on water with attractive mode (phase $>90\degree$) scan. Image size = $8 ~\mu m$.
(Bottom) Height (left) and phase (right) images of the same area with repulsive mode (phase $<90\degree$) scan.}
  \label{Sfig:afm}
\end{figure*}

\clearpage 

\begin{figure*}[t!]
\centering 
\renewcommand\thefigure{S\arabic{figure}}    
\includegraphics[width=\textwidth]{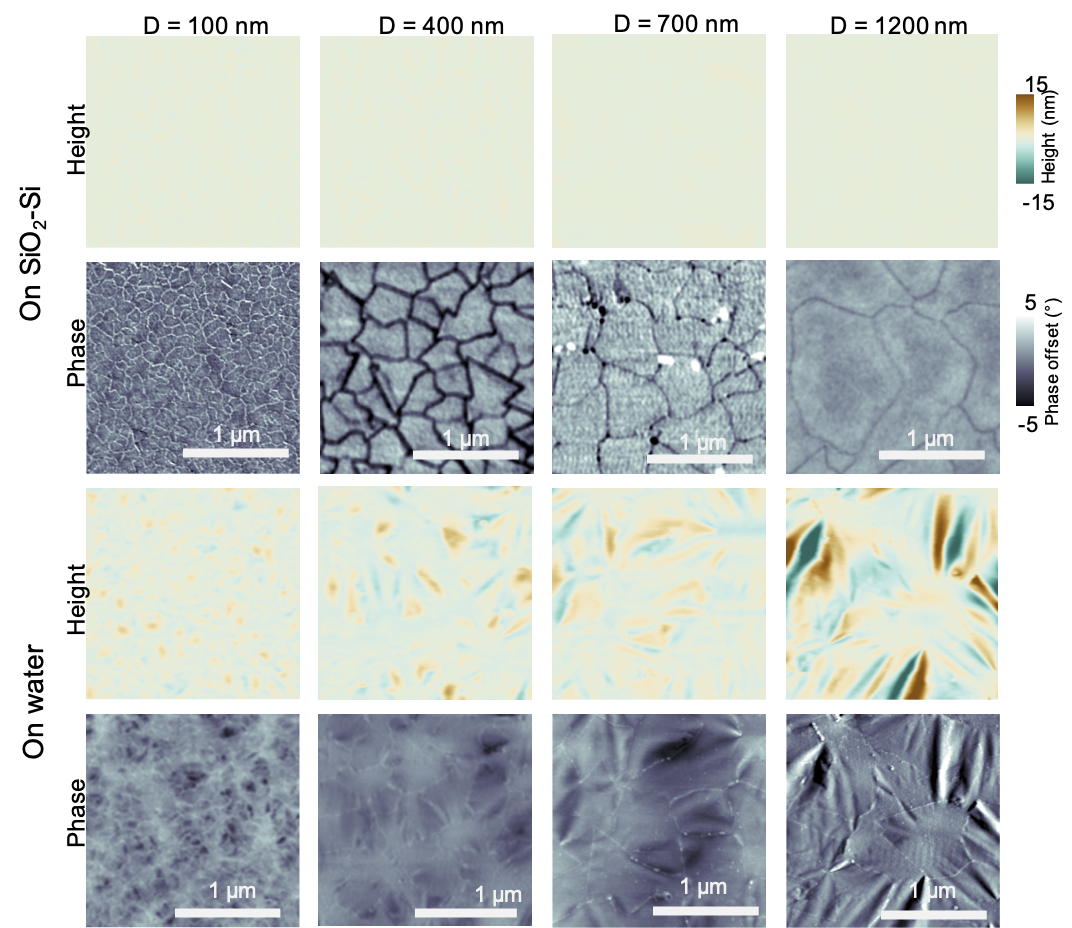}
 \caption{{\bf Height and phase images of \ch{MoS2} as grown on \ch{SiO2}-Si and on water.} (Top) Height and phase images from the monolayer \ch{MoS2} as grown with different domain sizes. The lines in phase images clearly show the grain boundary structures.
(Bottom) Height and phase images of \ch{MoS2} with varying domain sizes on a water surface.
  }
  \label{Sfig:heightphase}
\end{figure*}

\begin{figure*}[t!]
\centering 
\renewcommand\thefigure{S\arabic{figure}}    
\includegraphics[width=\textwidth]{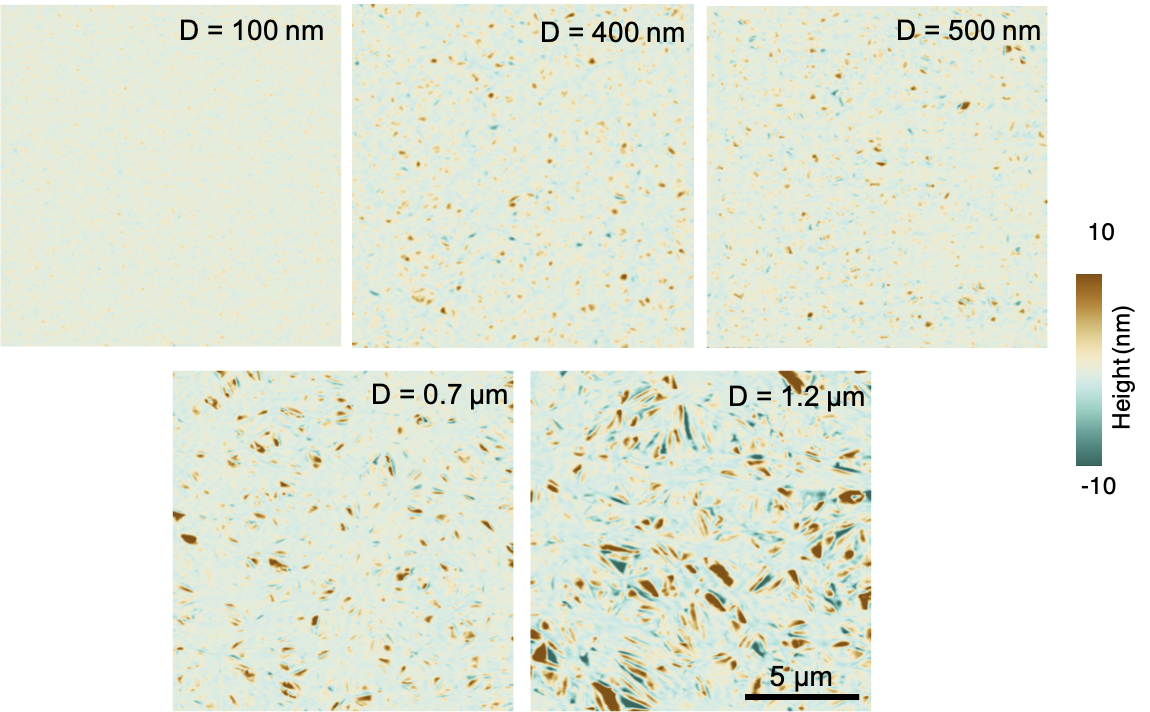}
 \caption{{\bf Grain-size dependent surface topographies of \ch{MoS2} floating on water.} 
The $15 \mu m \times 15\mu m$ surface topography scan of floating \ch{MoS2} with different domain sizes. }
  \label{Sfig:grainsize}
\end{figure*}

\clearpage

\begin{figure*}[t!]
\centering
\renewcommand\thefigure{S\arabic{figure}}    
\includegraphics[width=\textwidth]{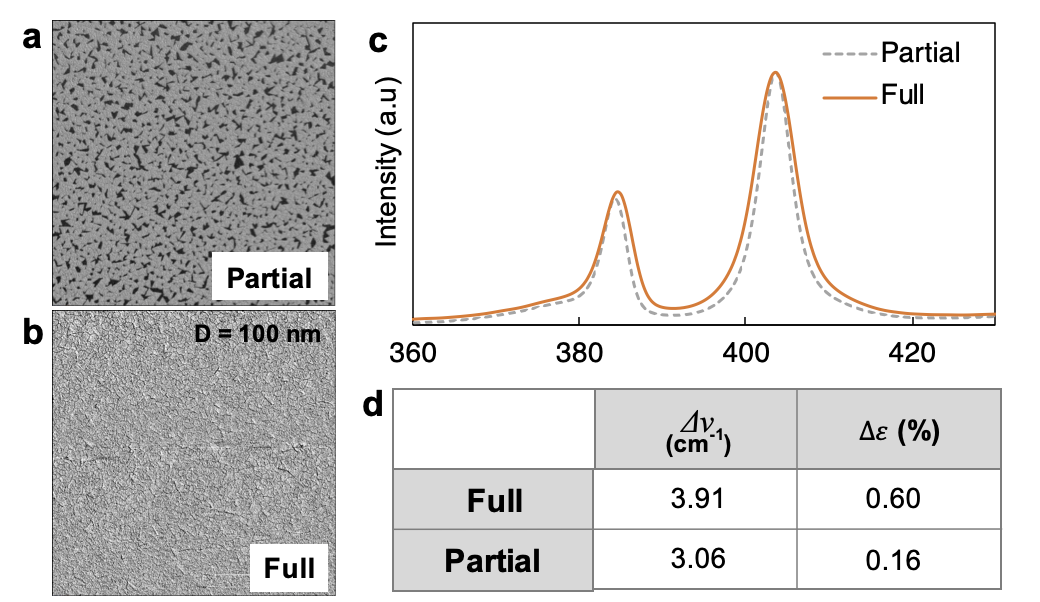}
 \caption{{\bf Formation of grain boundaries and strain distribution.} 
{\bf a-b.} Phase images of a monolayer \ch{MoS2}: {\bf a} before the formation of grain boundaries (Partial), and {\bf b} fully stitched (Full). The image size is $5 ~\mu m \times 5 ~\mu m$. The black area in (a) shows the bare \ch{SiO2}-Si surface.
{\bf c.} Average Raman spectrum obtained from the region shown in (a) and (b). The dotted line represents the spectrum of partial \ch{MoS2} (a), while the solid line represents that of full \ch{MoS2} (b).
{\bf d.} The Full Width at Half Maximum (FWHM) ($\Delta v_{cvd}$) of the $E_{2g}$ peaks in (c) and the corresponding strain variation.
  }
  \label{Sfig:formation}
\end{figure*}

\clearpage

\begin{figure*}[t!]
\centering 
\renewcommand\thefigure{S\arabic{figure}}    
\includegraphics[width=\textwidth]{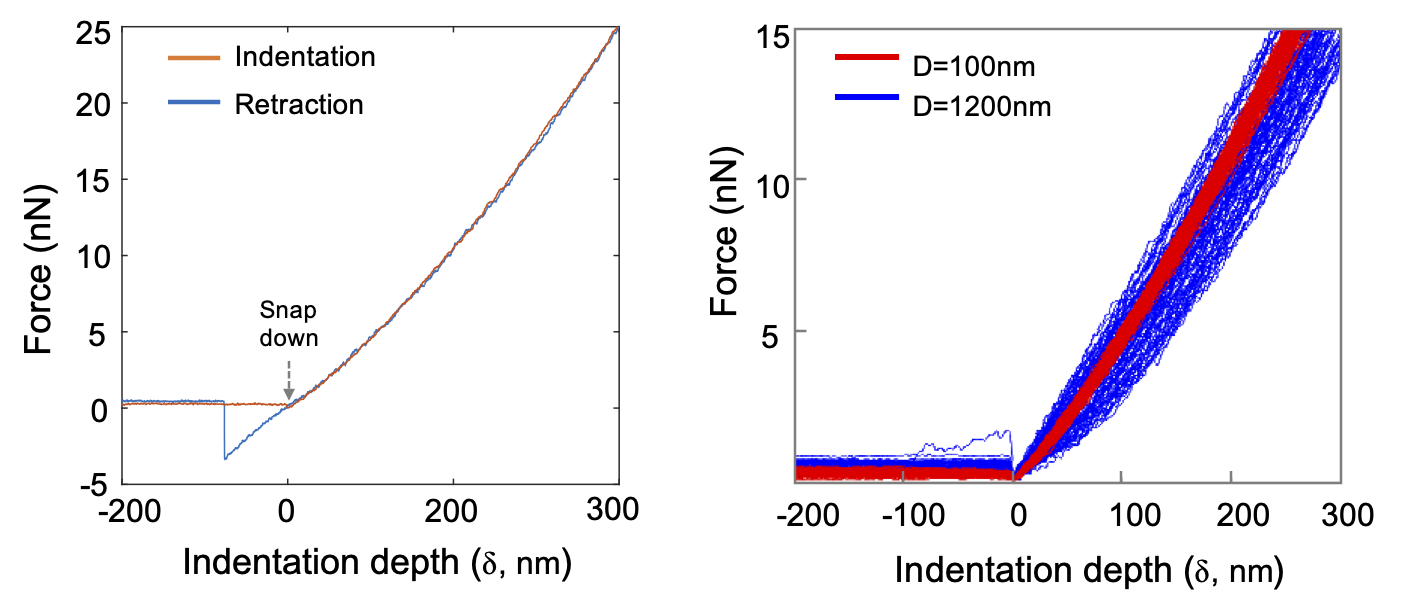}
 \caption{{\bf Force-indentation curve measurements.} 
(left) The force curves taken during a single indentation and retraction. (right) The indentation curves for small ($D = 0.1~\mu m$, red lines) and large ($D = 1.2~\mu m$, blue lines) grains for all locations in the force maps in Fig.~\ref{fig:fig4}d-e.}
  \label{Sfig:indentation}
\end{figure*}

\clearpage

\begin{figure}
    \centering
    \includegraphics[width=\textwidth]{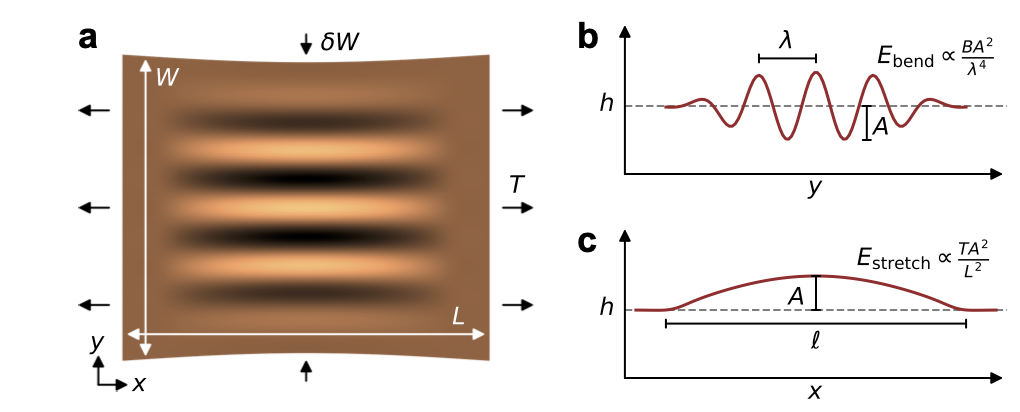}
    \caption{{\bf Tension-induced wrinkling schematic.}~{\bf a.} A schematic of a thin elastic sheet (dimension $L \times W$) undergoing wrinkling when compressed horizontally by a distance $\delta W$ and place under tension $T$. 
    {\bf b.}~A vertical cross-section of the height field $h(x=0,y)$ showing the amplitude $A$ and wrinkle width $\lambda$. The bending energy density of the wrinkles goes as the bending modulus times the square of the curvature: $E_\text{bend} \propto B (\nabla^2 h )^2  \propto B A^2/\lambda^4$.  
    {\bf c.}~A horizontal cross-section of the height field $h(x,y=0)$ showing the amplitude $A$ and the wrinkle length $\ell$. The stretching energy density is proportional to the tension times the change in arclength in the $y$-direction: $E_\text{stretch} \propto T (\partial_y h)^2 \propto T A^2/L^2 $. 
    }
    \label{fig:wrinkleschem}
\end{figure}

\end{document}